\documentclass[conference]{IEEEtran}

\newtheorem{definition}{Definition}

\usepackage{float}
\usepackage{seqsplit}
\usepackage{stfloats}
\usepackage[caption = false]{subfig}
\usepackage{listings}
\usepackage{xcolor}
\usepackage{multirow}
\usepackage{comment}

\usepackage{booktabs}

\usepackage[flushleft]{threeparttable}
\usepackage{adjustbox}
\usepackage{algpseudocode}

\usepackage{makecell}

\usepackage{enumitem}
\usepackage{soul}
\usepackage{algorithm}
\usepackage{xspace}
\usepackage{url}
\usepackage{pifont}
\newcommand{\cmark}{\ding{51}}%
\newcommand{\xmark}{\ding{55}}%

\newcommand{\new}[1]{\textcolor{black}{#1}}
\newcommand{\algoFullName}{Information-based Heavy Hitters\xspace}
\newcommand{\algoName}{\emph{ibHH}\xspace}

\begin{document}

\title{Information-Based Heavy Hitters for Real-Time DNS Data Exfiltration Detection and Prevention}

\setstcolor{red}

\author{\IEEEauthorblockN{Yarin Ozery}
\IEEEauthorblockA{Ben-Gurion University of the Negev\\ Akamai Technologies, inc. \\
yarinoz@post.bgu.ac.il}
\and
\IEEEauthorblockN{Asaf Nadler}
\IEEEauthorblockA{Ben-Gurion University of the Negev\\
asafnadl@post.bgu.ac.il}
\and
\IEEEauthorblockN{Asaf Shabtai}
\IEEEauthorblockA{Ben-Gurion University of the Negev\\
shabtaia@bgu.ac.il}}

\maketitle              


\begin{abstract}
Data exfiltration over the DNS protocol and its detection have been researched extensively in recent years. 
Prior studies focused on offline detection methods, which although capable of detecting attacks, allow a large amount of data to be exfiltrated before the attack is detected and dealt with. 
In this paper, we introduce \algoFullName{} (\algoName{}), a real-time detection method which is based on live estimations of the amount of information transmitted to registered domains.
\algoName{} uses constant-size memory and supports constant-time queries, which makes it suitable for deployment on recursive DNS servers to further reduce detection and response time.
In our evaluation, we compared the performance of the proposed method to that of leading state-of-the-art DNS exfiltration detection methods on real-world datasets comprising over 250 billion DNS queries. 
The evaluation demonstrates \algoName{}'s ability to successfully detect exfiltration rates as slow as 0.7B/s, with a false positive alert rate of less than $0.004$, with significantly lower resource consumption compared to other methods.
\end{abstract}

\section{\label{sec:intro}Introduction}

Data exfiltration is performed by malicious actors in order to steal data from a network. 
Once the data is collected, adversaries often utilize various techniques, including compression and encryption, to obfuscate the information and evade detection when removing it. 
The methods used to extract data from a targeted network typically involve transmitting it through a covert communication channel. 
Adversaries may also impose limitations on the data transmission size to minimize the risk of detection.
The data exfiltration stage usually marks the final phase in the malware lifecycle~\cite{strom2018mitre}.

The \emph{Domain Name System} (DNS) protocol~\cite{rfc1034} is a crucial network protocol, which is primarily used to translate easy-to-remember domain names into corresponding IP addresses. 
While enterprises and government organizations employ various defensive measures to safeguard their users and networks against malicious actors (e.g., antivirus software, firewalls, and network traffic monitoring), the DNS protocol is typically left unblocked and inadequately monitored due to its critical role in facilitating users' Internet access~\cite{lyu2022survey}.

Given the vulnerable and exploitable characteristics of the DNS protocol, malicious actors targeting enterprise and government organizations for data theft often choose to exploit it as a means of data exfiltration and covert communication.
This practice is known as \emph{DNS exfiltration}~\cite{bromberger2011dns}.

Initiating DNS exfiltration is a straightforward and cost-effective process, as the attacker simply needs to register a domain (e.g., \emph{attacker.com}) with a domain name registrar and assign an authoritative name server under their control to that domain.
This allows the malware installed on the compromised host to exfiltrate data by encoding it within DNS packets directed towards the registered domain. 
In the DNS query resolution process the queries are forwarded from the client to the authoritative name server associated with the queried domain. 
As a result, queries aimed at \emph{attacker.com} are forwarded to the attacker-controlled authoritative name server, enabling successful data exchange between the malware and the attacker. 
In addition, the attacker can utilize DNS responses to encode messages sent to the malware, enabling a bidirectional covert communication channel through the DNS exfiltration tunnel. 
This channel can be exploited for other purposes, such as command and control (C\&C) operations.

There are various ways to exfiltrate data over the DNS: 
(1) encoding data within the DNS query name (i.e., the target domain name to be resolved);
(2) using the DNS query type (qtype) field, which indicates the type of DNS resource record (RR) the client is trying to resolve, in order to encode a small amount of information (up to 16 bits) in the DNS packet;
and (3) timing-based exfiltration, in which the query arrival time is used as a means of conveying information~\cite{girling1987covert}.

Numerous instances of malicious actors leveraging DNS for data exfiltration have been documented, including activities by state-sponsored threat groups~\cite{denis,frameworkpos,oilrig-dns-tunneling,oilrig-fortinet,b1txor20}.
Moreover, the use of DNS for data exfiltration and communication has become increasingly prevalent among ransomware actors~\cite{dnsexfransomware}.
Therefore, it is no surprise that significant research attention has been dedicated to DNS exfiltration and its detection in recent years~\cite{wang2021comprehensive}.

Many DNS exfiltration methods have been proposed to date, ranging from rule-based techniques~\cite{paxson2013practical,ishikura2021dns}, statistical-based techniques~\cite{qi2013bigram,homem2018information}, supervised machine learning techniques~\cite{almusawi2018dns}, unsupervised machine learning techniques~\cite{nadler2019detection,ahmed2019real}, and even deep learning techniques~\cite{palau2020dns,chen2021dns,wu2020tdae}.

Despite the extensive body of research on DNS exfiltration, limited attention has been directed towards real-time detection methods. 
The majority of existing solutions are ill-suited for online deployment, and the proposed approaches have predominantly operated in an offline manner. 
Offline detection methods are a major concern as they allow for substantial data exfiltration to occur before the attack is identified and thwarted.

In contrast, real-time DNS exfiltration detection enables a rapid response from network operators, effectively reducing the potential damage inflicted by breaches. 
By quickly detecting and responding to DNS exfiltration attempts in real time, the negative consequences of such attacks can be mitigated.

In order to provide true real-time detection capabilities, a solution should not rely on an external data collection process but rather be executed directly on the DNS queries stream resolved by the resolver.
By integrating the detection functionality within the resolver itself, the solution can effectively analyze and classify queries without introducing delays or disruptions to the resolution process. 

Given that a true real-time detection solution should run directly on the resolver, it is crucial to ensure that the DNS resolution process remains unaffected by the detection mechanism. 
An effective solution should therefore possess low memory and computational demands while maintaining the ability to process and classify a large volume of queries per second.




In this paper, we introduce  \emph{\algoFullName}{} (\algoName), a novel and interpretable method for real-time DNS exfiltration detection. 
Our approach utilizes a threshold-based method to quantify the unique information transmitted through subdomains within DNS queries and raise alert if the suspected amount of data exfiltrated exceeds that threshold, making it transparent and explainable. 
To achieve true real-time detection, we employ a fixed-size data structure that efficiently processes a continuous stream of DNS queries, inspired by the concept of identifying heavy hitters in data streams~\cite{locher2011finding}.

\begin{figure*}[h]
    \centering
    \includegraphics[scale=0.5]{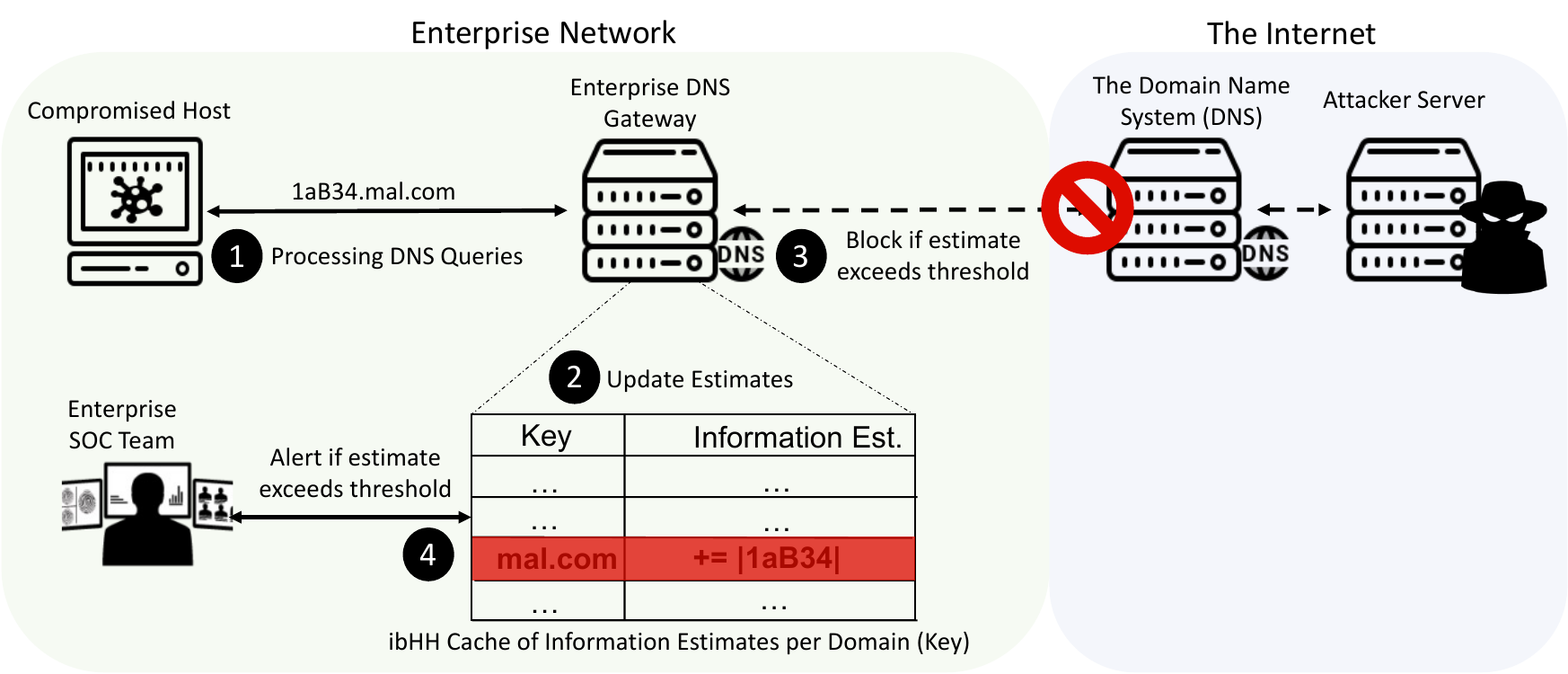}
    \caption{Overview of \algoName. 
    When a compromised host performs DNS queries (1) the enterprise DNS gateway intercepts the query, estimates the amount of information it contains and updates its internal information estimate cache (2).
    After the update, if the amount of information exceeds a predefined threshold, the query is blocked from reaching the attacker's server and (3) an alerted is raised for the enterprise's SOC team (4).}
    \label{fig:idhh_dns_algo}
\end{figure*}

\algoName{} incorporates the \emph{HyperLogLog} sketching algorithm~\cite{flajolet2007hyperloglog} from the field of big data~\cite{indyk2007sketching} and leverages weighted sampling techniques. 
This combination enables our method to accurately estimate the volume of information transmitted from clients to registered domains through subdomains. 
By comparing this estimated quantity against a predefined and configurable detection threshold, \algoName{} can raise an alert when the transmitted information surpasses the threshold. 
A detailed description of \algoName{} is provided in Section~\ref{sec:ibhhdefiniton}, and an overview of the method, along with a possible DNS exfiltration scenario, is presented in Figure~\ref{fig:idhh_dns_algo}.

In order to enable reproducibility of our results and allow further research, we provide an open-source Python implementation of our proposed solution.

Our experiments demonstrate \algoName{}'s high performance and its ability to process over 600,000 queries per second. 
This makes it well-suited for deployment in large-scale networks where real-time processing and performance are crucial. 
Additionally, \algoName{} maintains its efficacy in resource-constrained environments with limited computational and memory resources. 
Therefore, it can be effectively employed in small-scale environments without straining available resources.

An additional advantage of \algoName{} is that its operation does not rely on labeled training data. 
This characteristic makes the need for data annotation redundant and facilitates easier deployment and maintenance of the method.

To handle false positive cases, we propose two reputation-based allowlisting approaches, which are described in Section~\ref{sec:whitelist}.

We performed a thorough evaluation of \algoName{} to assess its effectiveness in detecting DNS exfiltration and its ability to handle false positive cases and compare it with three state-of-the-art (SOTA) methods proposed in prior studies.  

To ensure the reliability of our results, we collected a real-world DNS query dataset containing over 50 billion queries. 
This extensive dataset serves as a robust foundation for our evaluation, allowing us to make reliable assessments of \algoName{}'s performance.

In addition to the real-world dataset, we also performed the evaluation on a publicly available dataset.  This enables reproduction of our results and external validation of our findings. 

To further enhance the validity and robustness of our evaluation, we concluded the assessment by using a second real-world dataset.
This additional dataset consists of over \textit{250 billion queries}, collected over a period of three weeks. Our evaluation on this dataset is discussed in Section~\ref{sec:realworldeval}.
We also evaluated the resource utilization of the compared methods, specifically focusing on memory consumption and compute time, and the results of this evaluation are presented in Section~\ref{subsec:resource-usage}.

To the best of our knowledge, this evaluation is the most extensive and rigorous evaluation conducted in the field.

Detailed information about the evaluation process can be found in Section~\ref{sec:evaluation}, where we provide a comprehensive overview of our datasets and the methodology used to assess \algoName{}'s performance and the compared methods.

To summarize, the contribution of our work is:
\begin{enumerate}
\item \algoName{} - A lightweight and simple real-time DNS exfiltration detection method, which is appropriate for large-scale high throughput networks as well as small networks;
\item Evaluation of our method against state-of-the-art methods on large-scale real-world datasets as well as a publicly available dataset; 
\item Open-source Python implementation of our method, which will enable reproduction of our results and support further research.
\end{enumerate}

\raggedbottom

\section{Background \label{sec:background}}

\subsection{DNS stream model \label{subsec:model}}
We model DNS stream queries on the data stream model presented in~\cite{babcock2002models}:

\begin{definition}[Data Stream]\label{def:datastream}
A \textbf{Data Stream} $S$ is an ordered set of elements $x_1, x_2, ... x_n$ where each element is observed exactly once.
\end{definition}
Given definition ~\ref{def:datastream}, in the scope of this research, each element $x_i$ is comprised of a pair of values ($k_i$, $v_i$), where the values are taken from domains $K$ and $V$, respectively. $k_i$ is called the key of the pair, while $v_i$ is called the subkey of the pair.
More specifically, each element in the DNS stream is a DNS query's qname ~\cite{rfc1034}, extracted to $(domain, subdomain)$ pair, where $domain$ is the second-level domain (which we denote as primary domain, or just domain, for the rest of the paper) and $subdomain$ is the concatenation of the rest of the labels of the qname. For example, given the qname \textbf{a.b.example.com}, the domain is \textbf{example.com.}, the subdomain is \textbf{a.b}, and the DNS query stream element is (\textbf{example.com.}, \textbf{a.b}).

\subsection{Real-time DNS exfiltration detection \label{subsec:criteria}}
The DNS protocol has traditionally been low-demanding ~\cite{bindreq}; therefore, DNS resolvers are usually deployed on limited hardware ~\cite{dnsrequirements}.
Given that knowledge, we define criteria which must be satisfied in order for a DNS exfiltration detection algorithm to be considered appropriate for real-time detection:

  \begin{enumerate}[label=(\roman*)]
   \item Given a DNS stream of length $n$, the amount of space required should be sublinear with regard to $n$, i.e., have space complexity of $o(n)$. 
   \item The classification of a given DNS query should have time complexity of $\Theta(1)$.
  \end{enumerate}

These criteria are designed to ensure that a real-time DNS exfiltration detection solution can run on the DNS resolver server, without impacting the DNS resolution throughput of the server or have a large memory footprint, which is needed for the DNS protocol's caching mechanism ~\cite{rfc1034}.

\section{\label{sec:related}Related Work}
The topic of detecting data exfiltration over the DNS protocol has been the subject of nearly 30 recent studies~\cite{wang2021comprehensive}.


\textbf{Offline detection methods by design.} There is a wide variety of methods whose design limits them from being applied in real time.
The most notable design limitation is time-based aggregation feature extraction; for example, Nadler et al.~\cite{nadler2019detection} proposed an anomaly-based isolation forest~\cite{liu2008isolation} model to detect both high and low throughput DNS exfiltration, based on features extracted over a sliding window of size $\lambda$ hours; classification is done based on the latest $n_s$ windows, meaning up to $n_s*\lambda$ hours can pass by the the time of detection. 
This also means that at any given moment, the number of DNS queries that need to be stored in memory is $\Omega(\lambda \cdot n_s)$, and therefore it inherently cannot run in real time.
Ishikura et al.~\cite{ishikura2021dns} proposed a DNS exfiltration detection solution based on what they called cache-property-aware features. 
For each client in an enterprise network, they suggested maintaining a list of the last $n$ Fully Qualified Domain Names (FQDNs) the client accessed, in order to calculate a property called Access Miss Count, denoted as $AMC^t$, which indicates the number of FQDNs queried by the client in the last $t$ seconds. 
The authors proposed both a rule-based model and a long short-term memory (LSTM) based model to identify DNS exfiltration activity. 
The memory requirements of the solution grow linearly with the amount of clients in the network (both for the access list and the LSTM state of each client). 
Moreover, the solution does not identify which FQDN is suspected as being used for DNS exfiltration but rather only determines if exfiltration has occurred in a given time window (which ranged between 100 and 1,200 seconds in their experiments); this makes it difficult for security operations teams to identify the malicious domain.

Paxson et al.~\cite{paxson2013practical} presented an information-based method that provides an upper bound on the amount of information that can possibly be transmitted via DNS queries. 
The method groups DNS queries per primary domain and DNS source IP (referred to as ``client`` in their paper) on a daily basis, which is followed by lossless compression of the different possible information vectors (query name, query timing, and query type); the minimal value of these is output as the upper bound on the amount of information. 
The upper bound is then compared with a predefined threshold, and alerts are raised for any communication which exceeds this value. 
This method is designed to run on window-aggregated data of size $w$ and has the benefit of being able to detect DNS exfiltration regardless of the information vector. 
However, both the time and space complexity of the method are $\Omega(w)$, because all of the queries in the window $w$ need to be kept for the information estimation stage, and the time complexity is $\Omega(w)$ due to the compression performed to estimate the information.




\textbf{Compute-intensive detection methods.} In recent years, deep learning-based (DL) DNS exfiltration detection methods have been proposed~\cite{zhang2019dns,palau2020dns,chen2021dns}.
Chen et al.~\cite{chen2021dns} proposed a DL architecture based on the combination of a convolutional neural network (CNN) and Long short-term memory networks (LSTM). 
For the LSTM layer, the authors used one hidden layer with length 128, as they assume that the first 128 characters of the DNS query may be used for exfiltration. 
Three convolution layers, two max-pooling layers, and one softmax layer were used for the CNN layer. Their model is then trained on labeled DNS queries (benign and malicious).
\new{Wu et al.~\cite{wu2020tdae} proposed TDAE, an autoencoder DL DNS exfiltration detection method based on semi-supervised learning, which means it requires some labeled data. }
While DL models generally provide high accuracy and automatic feature extraction, they come with the cost of requiring larger training datasets than traditional machine learning methods.
More importantly, they are known to be hardware-intensive~\cite{luong2019applications,mohammed2019machine}, which makes them unsuitable for deployment on the network perimeter. 


\textbf{Supervised learning methods.} Several methods that rely on labeled data for training~\cite{almusawi2018dns,buczak2016detection,yang2020naruto,ruiling2022dns} have been proposed. 
This is reasonable for identifying a predetermined set of known DNS exfiltration tools, but as shown in~\cite{nadler2019detection}, the absence of high-quality publicly-available datasets prevents these methods from identifying unfamiliar DNS exfiltration malware.
Our proposed method does not require any labeled data for training.

\textbf{Real-time detection methods.} 
To the best of our knowledge, only two previous studies focused on real-time detection~\cite{qi2013bigram, ahmed2019real}.
Qi et al.~\cite{qi2013bigram} suggested a detection technique based on bigram (subsequent pairs of characters) frequency statistics. 
The authors described a score mechanism based on the expected value of the bi-gram character frequency as the score of the primary domain. 
The model is trained on labeled benign and malicious data to determine the score threshold that will make the classifier produce the least number of false positive alerts, which is then used by the classifier in the online phase. 
While it is reasonable to expect this kind of classifier to be run in real time, it suffers from the same limitation of other supervised learning methods -- it has difficulty generalizing to unfamiliar DNS exfiltration techniques.
Ahmed et al.~\cite{ahmed2019real} proposed an unsupervised isolation forest model, which is based on classifying queries on a per packet basis. 
Many features are extracted and used, such as the length of the query name, length of the query subdomain, query name entropy~\cite{shannon1948mathematical}, count of numerical characters in the query name, count of uppercase letters in the query name, number of DNS query labels, maximum label length, and average label length. 
While the isolation forest model~\cite{liu2008isolation} is relatively lightweight in terms of classification time and memory requirements, it is questionable how this method can scale to large-scale networks which may reach millions of queries per second~\cite{schomp2020akamai}, given the number of features that need to be extracted to perform classification. 
A table summarizing related studies and their methods' compliance with the real-time criteria defined in Subsection~\ref{subsec:criteria} is provided in Table~\ref{tab:relatedworkstable}.



\begin{table}[h]
\centering
\begin{adjustbox}{width=\columnwidth}
\begin{threeparttable}
\begin{tabular}{l|c|c|c|c|c|cc}
\hline
Study &
  Year &
  Technique &
  \makecell{Requires \\ Labeled \\ Training \\ Data} &
  \makecell{Requires \\ Data \\ Aggregation} &
  \makecell{Classification \\ in Constant Time} &
  \makecell{Suitable \\ For Real-Time \\ Detection} \\
  \hline
Paxson et al. ~\cite{paxson2013practical} &
  2013 &
  $RB$ &
  \xmark{} &
  \cmark{} &
  \xmark{} &
  \xmark{} \\
  \hline
 Kara et al. ~\cite{kara2014detection}    & 2014 & $RB$ & \xmark{}   & \cmark{}   & \xmark{}   & \xmark{}   \\
 \hline
 Buckzak et al. ~\cite{buczak2016detection} & 2016 & $MB$ & \cmark{}   & \cmark{}   & \xmark{} & \xmark{}   \\
 \hline
 Almusawi et al. ~\cite{almusawi2018dns} & 2018 & $MB$ & \cmark{}   & \xmark{}   & \xmark{}   & \xmark{}   \\
 \hline
 \new{Homem et al.} ~\cite{homem2018information} & 2018 & $MB$ & \xmark{}   & \cmark{}   & \cmark{}   & \xmark{}   \\
 \hline
 Nadler et al. ~\cite{nadler2019detection}   & 2019 & $MB$ & \xmark   & \cmark   & \xmark{} & \xmark   \\
 \hline
 Ahmed et al. ~\cite{ahmed2019real}    & 2019 & $MB$ & \xmark   & \xmark   & \cmark   & \cmark   \\
 \hline
 Palau et al. ~\cite{palau2020dns}    & 2020 & $MB$ & \cmark   & \xmark   & \xmark   & \xmark   \\
 \hline
  \new{Wu et al. ~\cite{wu2020tdae} }   & 2020 & $MB$ & \cmark   & \xmark   & \xmark   & \xmark   \\
 \hline
  \new{Yang et al. ~\cite{yang2020naruto}}   & 2020 & $MB$ & \cmark   & \xmark   & \xmark   & \xmark   \\
 \hline
  \new{Ishikura et al. ~\cite{ishikura2021dns}}    & 2021 & $MB/RB$ & \xmark   & \cmark   & \xmark   & \xmark   \\
 \hline
  \new{Ruiling et al. ~\cite{ruiling2022dns}}    & 2022 & $MB$ & \cmark   & \xmark   & \xmark   & \xmark   \\
 \hline
 \algoName      & 2022 & $RB$ & \xmark   & \xmark   & \new{\cmark}   & \cmark \\
 \hline
\end{tabular}
\end{threeparttable}
\end{adjustbox}
\caption{Comparison of previously proposed methods for detecting DNS exfiltration (RB-rule-based, MB - model-based) }
\label{tab:relatedworkstable}
\small
\end{table}

\section{Information-Based Heavy Hitters for DNS Exfiltration Detection}

\subsection{Definitions}

\begin{definition}[Information Weight]\label{def:informationweight}
Given a stream of elements $S$, the information weight of an element ($k_i, v_i$), $I_{k_i, v_i}$ is the quantity of information conveyed by $v_i$.

\end{definition}

\begin{definition}[Distinct Information Heavy Hitter]\label{def:distinctinformationheavyhitter}
In a stream of elements $S$, for a given $k_i \in K$,
the distinct information weight $I_{k_i}$ is the total information conveyed by distinct  elements with key $k_i$, i.e., \[
   I_{k_i} = \sum_{ \{v |(k_i,v)\in S \}} I_{k_i,v}
\] \\ Key $k_i$ is a \textbf{distinct information heavy hitter} if its information weight $I_{k_i}$ is at least $\epsilon$-fraction of the total distinct information weight of the stream, where $\epsilon \in (0,1)$, i.e.,  \[
   I_{k_i} \geq \epsilon \cdot \sum_{y\in K} I_y
\] \\
\end{definition}

\subsection{\algoFullName \label{sec:ibhhdefiniton}}
\algoFullName (\algoName) is a novel method for real-time detection of DNS exfiltration, which is based on identifying domains associated with a large amount of distinct information conveyed through subdomains in a DNS query stream and inspired by the work of Afek et al.~\cite{afek2016efficient} in which heavy hitter detection algorithms were proposed for the detection of DDoS attacks.
Essentially, the task of identifying heavy hitters in a data stream is to find the most frequent keys in a stream, while the purpose of identifying distinct heavy hitters is to find keys that appear with the most distinct subkeys in a stream.  
Despite the usefulness of solving these two problems for various tasks, such as DDoS detection~\cite{afek2016efficient} and traffic load balancing~\cite{yang2016heavy}, they do not fully capture the complexity of DNS exfiltration detection, where there is a need to account for the amount data being exfiltrated via DNS queries. 
In order to model this complexity, we introduce the new concepts of \emph{information weight} (see Definition~\ref{def:informationweight}) and \emph{information-based distinct heavy hitters} (see Definition~\ref{def:distinctinformationheavyhitter}), which describe elements associated with large amounts of unique information in a stream.
\algoName{} quantifies the amount of information conveyed through DNS query subdomains to domains, identifies domains associated with large amounts of unique information, and raises alerts for these domains as suspected for DNS exfiltration.
The input for \algoName{} is a stream of DNS queries, such that for each DNS query \emph{subdomain.example.com}, the domain and subdomain are extracted to obtain the element: $(example.com, subdomain)$. \algoName{} consists of a fixed-size cache ($Counters$) whose size ($k$) is a parameter of the algorithm; a random hash function $Hash ~ U[0,1]$ that allows us to sample the distinct DNS query stream; $detection\_threshold$, which is a parameter of the algorithm; and a threshold value $\tau$ (initialized to $1$), which represents the probability of a domain's inclusion in the cache.

Each entry in the cache stores an information counter, which calculates the total unique information weight for $domain$, i.e., $I_{domain}$  and $seed_{domain}$, whose value is the minimum \\ $Hash(domain,subdomain)$ of all elements with key $domain$ in the stream.
\subsubsection{Information Quantification\label{subsec:infoquant}}
According to Definition \ref{def:distinctinformationheavyhitter}, we need to quantify the amount of unique information encoded in subkeys and calculate this amount per key.  
In order to do so, we need a function $I: V \rightarrow I_V$, where $I(subdomain)$ is the information weight of $subdomain$. 
There are different ways that the information weight can be defined, but in the scope of this research, we denote $I(subdomain) = |subdomain| $, i.e., the information weight of a subkey is its length. 
We acknowledge that this choice does not provide an exact quantification of the information encoded in the subdomain, but it imposes an upper bound on the quantity of information that can be conveyed through it, and as will be shown in our experiments in Section \ref{sec:evaluation}, it provides an adequate approximation for practical purposes. 
We also experimented with using entropy~\cite{shannon1948mathematical} to estimate the information, but the results were inferior.

%
\subsubsection{Optimized Counting with HLL++\label{subsec:optimizecount}.}
In order to calculate the exact amount of unique information, for each domain we need to store a set of all its associated subdomains in a stream, which requires linear space complexity.
Since our solution is intended to run on DNS resolvers with limited memory capabilities, we cannot use exact information weight counters. Instead, we employ count-distinct approximation algorithms from the world of big data.
The count-distinct problem is a well-studied problem~\cite{rajaraman2014mining}, and many extremely accurate and high-performing approximation algorithms exist for it. 
One of the state-of-the-art solutions for this problem is \emph{HyperLogLog} (\emph{HLL})~\cite{flajolet2007hyperloglog}. 

Essentially, HyperLogLog takes advantage of a clever property of multisets that the cardinality of a multiset (the number of distinct elements) of uniformly distributed random numbers can be estimated by calculating the maximum number of leading zeros in the binary representation of each number in the set. 
If the maximum number of leading zeros observed is $l$, then the number of distinct elements in the multiset is approximately $2^l$. 
In order to obtain a uniformly distributed random number multiset, a hash function is applied to the elements in the original multiset, which is the DNS stream in the scope of this paper. 
\emph{HLL}'s data is stored in counter arrays, which are called registers, and the size of the arrays depends on the number of bits allocated for the registers, $p$. In this research, we fixed $p$ at 12, in order to achieve optimal memory consumption while achieving highly accurate cardinality approximations.
In the proposed method, we use a variation of the original  \emph{HLL}, known as \emph{HyperLogLog++} (\emph{HLL++})~\cite{heule2013hyperloglog}, which provides better accuracy and uses less memory than the original design. 
For each cached domain, we store an instance of \emph{HLL++} which is used to approximate the total distinct information weight of subdomains of DNS queries.
\emph{HLL++} approximates the distinct number of elements in a stream, while our intention is to approximate the amount of distinct information in a stream, we manipulate the input for \emph{HLL++}, such that for each DNS query stream element $(domain, subdomain)$, instead of adding $subdomain$ to $domain's$ \emph{HLL++} instance, for each integer $i$ in the range of $(0, length(subdomain))$, we add the concatenated string $subdomain || i$ to $domain's$ \emph{HLL++} instance; thus, we are able to approximate the amount of information conveyed to $domain$. If the value of the approximation information counter for $domain$ exceeds $detection\_threshold$, an alert is raised.

When a $(dom, sub)$ element is processed, we check whether $dom$ is in $Counters$; If $dom$ is not in the $Counters$ cache, we calculate the value $h = Hash(dom, sub)$.
Then, if $h<\tau$, we initialize the $HLL++$ instance for $dom$, increase its information counter appropriately, and initialize a $seed$ value for it, denoted as $seed_{dom}$, to be $h$.
Otherwise, we insert the element by inserting it into the \emph{HLL++} instance above, calculate $ h = Hash(dom, sub)$, and $seed_{dom}$ is updated to be the minimum between $seed_{dom}$ and $h$.

If the size of $Counters$ exceeds $k$ after the insertion of $dom$, we evict the domain with the largest $seed$ value from the cache, and $\tau$ is updated to be the evicted key's $seed$ value; thus $\tau$ acts as the probability of being included in the cache sample. This idea was first introduced in the work of Gibbons et al~\cite{gibbons1998new}. 

By updating the $seed_{dom}$ value whenever we process a pair with $dom$ as the key and considering that the value can only decrease, it suggests that the evicted domain is likely to be the least information-heavy hitter. 
This approach increases the probability that the cache primarily contains the most significant domains in terms of information volume.

\algoName's pseudocode  is provided in Algorithm \ref{alg:idHH}, and an overview of \algoName is presented in Figure \ref{fig:idhh_dns_algo}.

\begin{algorithm}
\caption{\algoName pseudocode}\label{alg:idHH}
\hspace*{\algorithmicindent} \textbf{Input}  $k$ - cache size, $d$ - detection threshold, stream of DNS queries
\begin{algorithmic}
\small
\State $\tau \gets 1$
\State $Counters \gets \{\}$
\For{\texttt{ stream element (domain, subdomain)}}
    \State $N \gets |subdomain|$
    \State $h \gets Hash(domain, subdomain)$
    \If{ domain is in Counters}
        \For{\texttt{i=0; i<N; i++}}
            \State $subdomain_i \gets subdomain || str(i)$
            \State $Counters[domain].add(subdomain_i)$
        \EndFor
        \State $seed_{domain} \gets min\{ seed_{domain}, h \}$
        \If{ $Counters[domain].InformationEst > d$}
            \State raise alert for $domain$
        \EndIf
    \Else
        \If{ $h < \tau$}
            \State $Counters[domain] \gets new  HLLPlusPlus$
            \State $seed_{domain} \gets h$
            \For{\texttt{i=0; i<N; i++}}
                \State $subdomain_i \gets subdomain || str(i)$
                \State $Counters[domain].add(subdomain_i)$
            \EndFor
            \If{ $|Counters| > k$}
                \State $ToDel \gets argmax_{dom \in Counters} seed_{dom}$
                \State $\tau \gets seed_{ToDel}$
                \State Delete $Counters[toDel]$
                \State Delete $seed_{ToDel}$
            \EndIf
        \EndIf
    \EndIf
\EndFor
\end{algorithmic}
\end{algorithm}

\subsection{\algoName{} Space and time complexity analysis \label{subsec:timespaceanalysis}}

One of \algoName{}'s benefits is the fact that it has sublinear (in fact, logarithmic) space complexity in the DNS stream length $n$, as well as constant query classification time complexity, which means that it satisfies the real-time criteria defined in Section~\ref{subsec:criteria}:

\subsubsection{Memory Analysis}

For our cache structure, we store k \emph{HLL++} instances. 
Using the $(\epsilon, \delta)$ model~\cite{heule2013hyperloglog}, each \emph{HLL++} instance requires $\mathcal{O}(\epsilon^{-2}loglog(m_{dom})+log(m_{dom}))$ space~\cite{heule2013hyperloglog}, where $m_{dom}$ is the number of distinct elements associated with key $dom$. 
We denote $m = max_{ dom \in Counters} m_{dom}$ for readability; thus the total space complexity of \algoName{} is $\mathcal{O}(k \cdot \epsilon^{-2}loglog(m_)+log(m))$, which is logarithmic in the cardinality of the data stream and therefore logarithmic in the entire data stream size, given that $m=O(n)$.

\subsubsection{Time Analysis (processing a query)}

When processing a DNS stream element (domain, subdomain), we calculate the hash value $h=Hash(domain, subdomain)$, which has a time complexity of $O(1)$. 
If \textbf{domain} is already cached, we proceed with adding \textbf{subdomain} to \textbf{domain}'s HLL++ instance. 
The add operation of the HLL++ algorithm has a time complexity of $O(1)$, and given that a domain name is limited to 255 characters, as described in the original DNS RFC~\cite{rfc1034}, the subdomain is also necessarily limited to 255 characters; therefore the time complexity of the add operation is $O(1)$. 
Following the add operation, \algoName{} performs a $count$ operation to determine if an alert should be raised for the domain. 
The $count$ operation has a time complexity that depends on the number of register bits allocated to the \emph{HLL++} instance $p$. 
We fix $p$ at $12$; therefore this operation has a time complexity of $O(1)$. 
We conclude that processing a query and classifying it has a constant time complexity.

\subsection{Reset mechanism \label{subsec:resetmechanism}}

Given the conditional probabilistic nature of the method (which derives from the need to calculate a hash function and compare it to the value of $\tau$), domains that appear earlier in the stream are much more likely to be included than later ones (due to the decreasing nature of $\tau$); thus the confidence interval of the counters decrease over time. 

In order to avoid missing information heavy hitters that appear later in the stream, we propose a reset mechanism, where the \algoName{} cache is flushed and reset at constant intervals, which allows us to identify DNS exfiltration events that occur long after the deployment of the \algoName{} algorithm in the network. 

\subsection{Allowlists \label{sec:whitelist}}

Patterns of legitimate use resembling that of DNS exfiltration makes it difficult to distinguish between benign and malicious DNS traffic. 
Anti-malware agents are known to use the DNS protocol to send signatures of suspected files to the DNS zone of the anti-malware service provider for inspection~\cite{nadler2022vulnerability}. 
Other services, such as search engines, social networks, and streaming services are known to use disposable domains for purposes of signaling~\cite{chen2014dns,zeng2021finding}. 
In order to handle false positive domains and avoid raising many false alarms, we present two simple lightweight allowlisting approaches based on the concept of global and local reputation, as described in~\cite{hu2016baywatch}. 
Globally reputable domains are domains listed in publicly available lists associated with benign domains (such as TRANCO) and therefore should be trusted.
Locally reputable domains, are domains queried by a large portion of hosts in the local enterprise network and therefore should be trusted. 

\new{Using the allowlists in a pre-filtering phase is also important for preventing known benign domains that have many distinct subdomains from being cashed and ensuring that they do not take up cache space.
In our evaluation, we apply the allowlists as a post-filter (instead of a pre-filter), in order to evaluate the effectiveness of the proposed approaches on the false positive rate of \algoName{} and the compared methods.}


\subsubsection{TRANCO\label{sec: tranco}}

The use of top-ranking domain lists for allowlist purposes is very common in DNS exfiltration detection~\cite{ahmed2019real,nadler2019detection,paxson2013practical}. 
In this paper, we use TRANCO~\cite{pochat2018tranco}, an approach for ranking websites' popularity, to generate a top 1M list that allows us to filter out popular websites, reducing the number of false positive domains.

\begin{table}[]
\adjustbox{width=\columnwidth}{

    \centering
    \begin{tabular}{c|c|c|c|c}
    \hline 
      Dataset Name & \# DNS Queries & \# Unique 2LD & \# Org & \# Hosts  \\
    \hline
    $DS_f$ & 50,853,030,033 & 43,310,209 & 753 & Unknown \\
    \hline
     $DS_p$ & 5,069,006,334  & 668,456 & 223  & 129,528 \\
     \hline
     \new{Ziza et al. ~\cite{Ziza2022DNSExfiltrationDataset}} & \new{35,074,149} & \new{12,844} & \new{N/A} & \new{35,989} \\
     \hline
     $DS_r$ & \new{255,750,980,779} & \new{463,122,409} & \new{753} & \new{Unknown} \\
     \hline
    \end{tabular}
    }
    \caption{A summary of the datasets used in this study}
    \label{tab:dataset_table}
\end{table}

\subsubsection{Peacetime/Wartime \label{sec:ptwt}}
Using the peacetime/wartime model, which was first introduced in~\cite{afek2016efficient}, we execute \algoName{} in a non-enforcing mode for a limited period of time. During this execution (called \emph{peacetime}), we assume that the presence of DNS exfiltration traffic in the network is negligible (inspired by the idea of~\cite{nadler2019detection}), and therefore any domain that is detected as malicious by \algoName{} during this time is actually benign and should be allowlisted. We collect these domains in an allowlist called the \emph{peacetime allow list}.
After that, \algoName{} is deployed in an enforcing mode (called \emph{Wartime}), filtering out domains that appear in the peacetime allowlist. 
This approach is model-agnostic; therefore we use it for \algoName{} and the compared methods in our evaluation. 
As will be shown, this approach is simple yet very effective.
The amount of time the algorithm should run in \emph{peacetime} mode depends on the network. 
We found that even an hour's worth of data results in a useful allowlist; therefore, we recommend running it for at least an hour.

\begin{figure}[h]
    \centering
    \subfloat[False positive queries]{\includegraphics[width = 0.8\columnwidth]{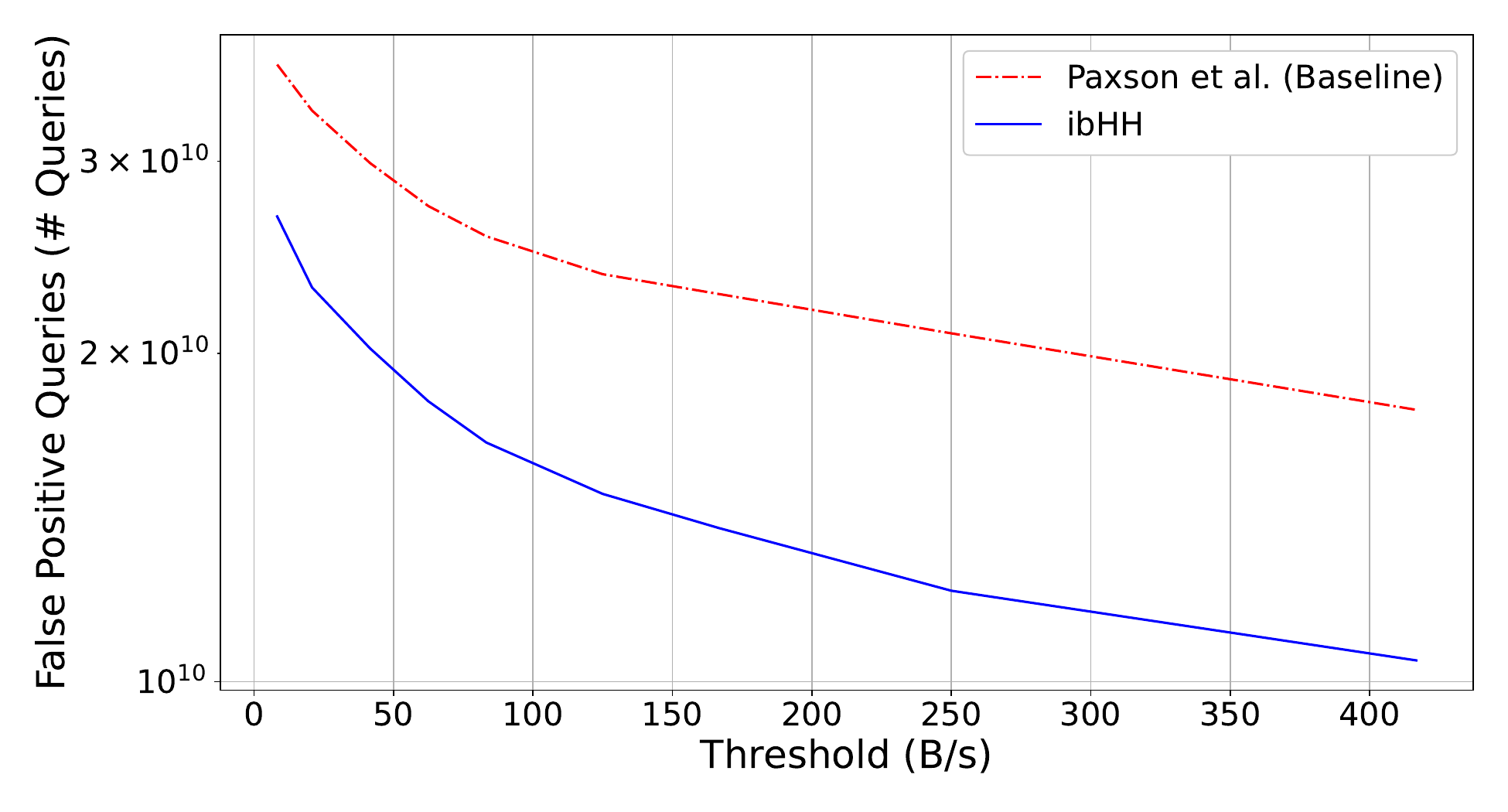}} \\
    \subfloat[False positive domains]{\includegraphics[width = 0.8\columnwidth]{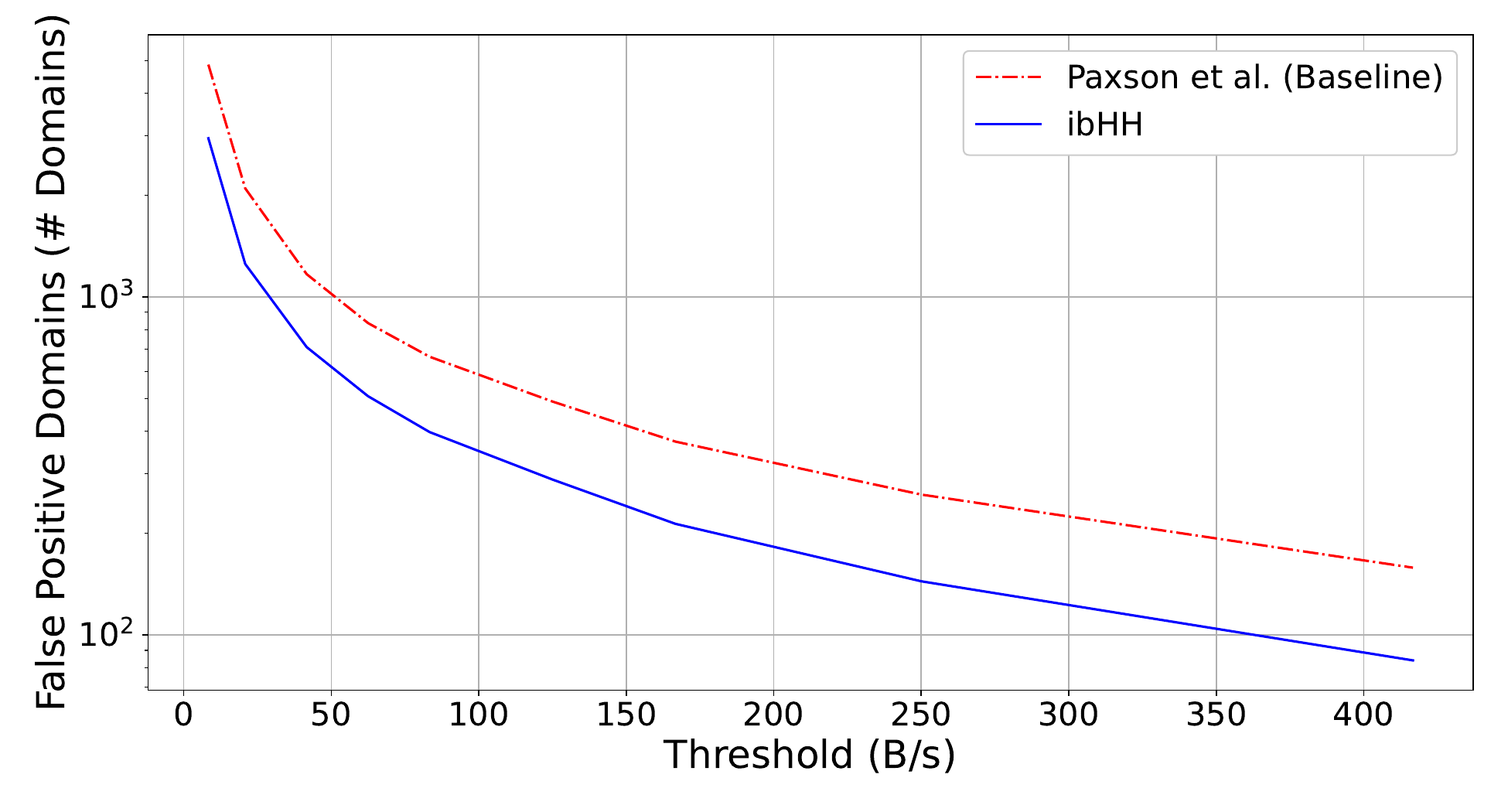}}
    \caption{Parameter tuning without allowlists.}
    \label{fig:global-fp}
\end{figure}

\section{Evaluation \label{sec:evaluation}}

\subsection{Overview \label{subsec:eval_overview}}
The evaluation is divided into two parts, namely parameter tuning and comparison with other methods. 

\textbf{Parameter tuning.} We compare the effect of different detection threshold values on the number of alerted queries and domains, as well as the effect of employing the proposed allowlist techniques.
The method of Paxson et al. is tuned similarly and serves as a baseline.
\begin{figure}[h]
    \centering
    \subfloat[False positive queries]{\includegraphics[width = 0.8\columnwidth]{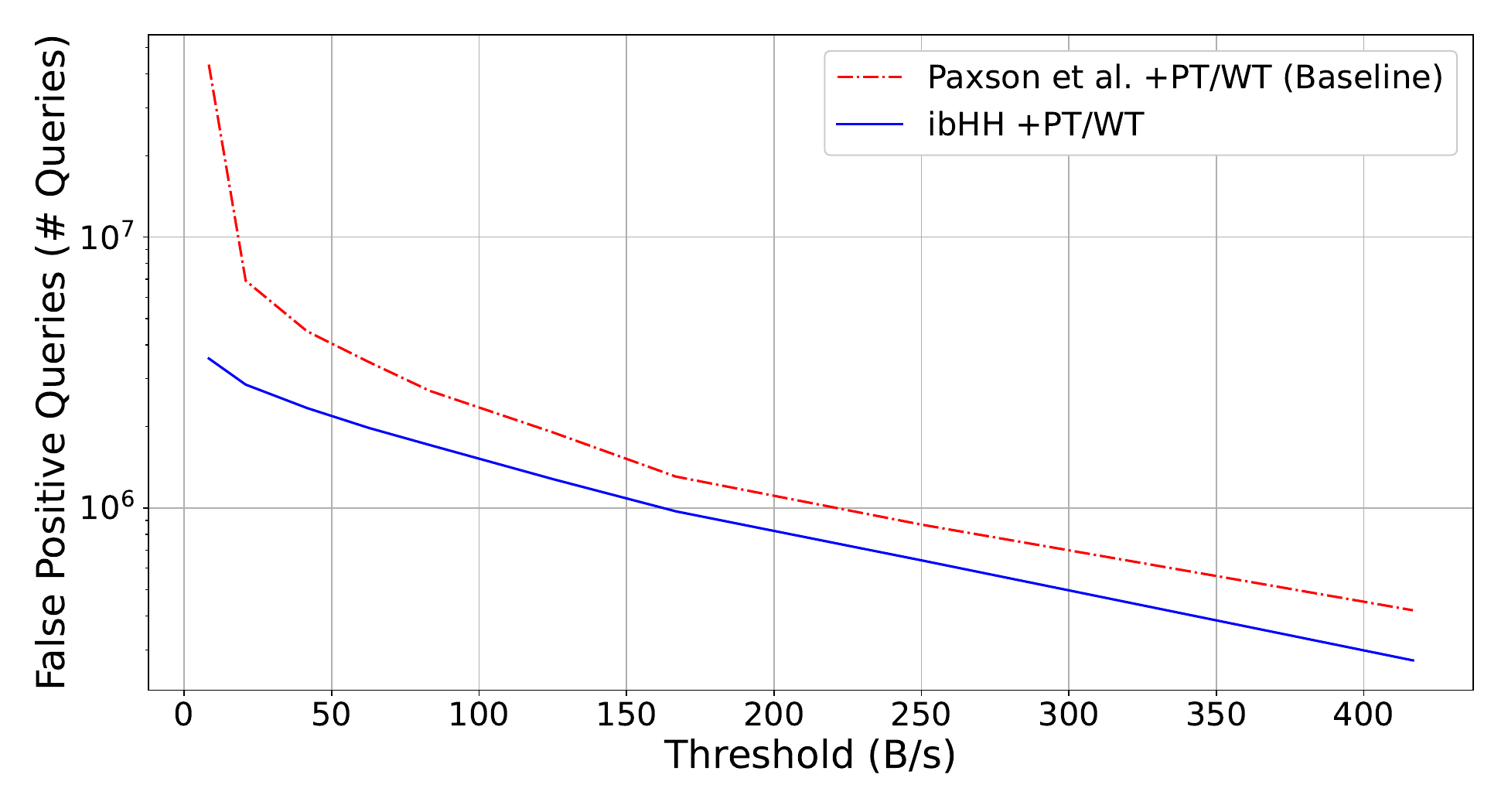}} 
    \\
    \subfloat[False positive domains]{\includegraphics[width = 0.8\columnwidth]{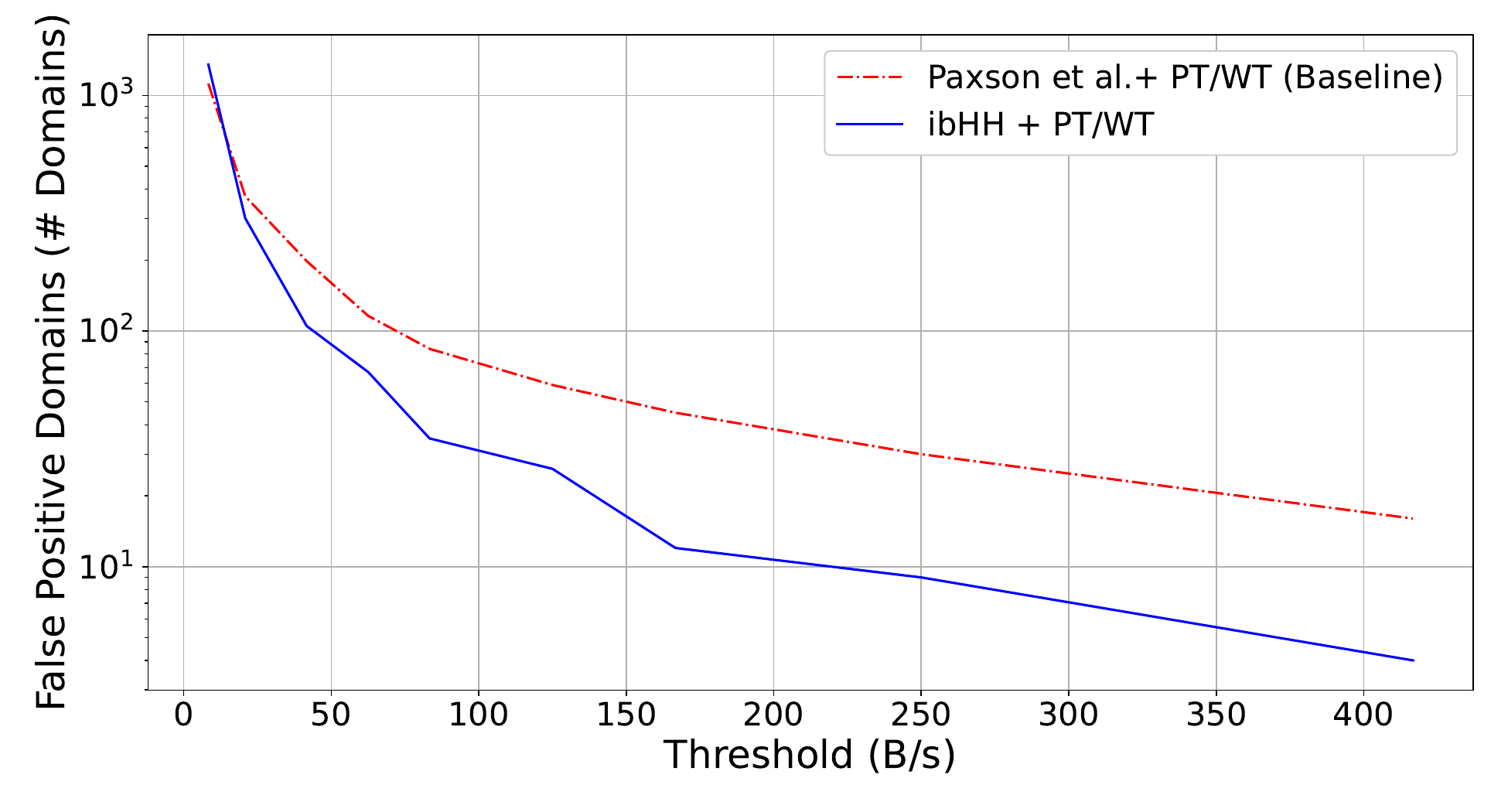}}
    
    \caption{Parameter tuning with peacetime allowlist}
    \label{fig:global-fp-pt}
\end{figure}
In addition, we present two deployment settings for \algoName{}; one simulates consolidation of data to a single point and its analysis (similar to other offline methods), and the other is a deployment setting that simulates \algoName{}'s execution right on the enterprise DNS gateway, and compare their performance.

\textbf{Comparison with other methods.} In this step, we evaluate our method's detection capabilities and compare it with the capabilities of methods proposed in earlier studies, namely the studies of Paxson et al.~\cite{paxson2013practical}, Nadler et al.~\cite{nadler2019detection}, and Ahmed et al.~\cite{ahmed2019real}. 
Our evaluation includes both detection efficacy comparison (the ability to properly identify DNS exfiltration domains and avoid misclassifcation of benign DNS domains), as well as a performance evaluation, comparing the classification time and memory use of the compared methods. 




\subsection{Datasets \label{subsec:dataset}}

The DNS traffic datasets used in this study were collected from DNS servers operated by a large CDN (content delivery network) provider. \new{In addition, a publicly available dataset published by Ziza et al. in November 2022~\cite{Ziza2022DNSExfiltrationDataset} is used as an independent dataset.}

\textbf{First dataset.} The first dataset, denoted as $DS_f$, consists of 50.85 billion DNS queries from 753 real-world enterprise organizations whose traffic is monitored by the CDN provider.
The queries were collected over the course of eight days, beginning on December 28, 2021.
Accordingly, the average number of queries per hour is 260 million.
The dataset contains one domain suspected of DNS exfiltration (as alerted by the CDN provider's proprietary DNS exfiltration algorithm), \url{joinsanjose[.]com}. 
Due to the scarcity of data exfiltration events and the fact that the dataset consists of monitored data, we presume that the rest of the dataset has at most just a negligible amount of malicious traffic, and we thus treat it as benign. 

\textbf{Identifiable dataset.} We sample a subset of $DS_f$, denoted as $DS_p$, which contains 5.06 billion DNS queries.
In contrast to the full dataset, all of the DNS queries in $DS_p$ can be attributed to a specific host of nearly 130,000 \emph{specific} IP addresses. 
The IP addresses are hashed to preserve the privacy of the end users.
The $DS_p$ subset is generated to evaluate the detection methods' abilities to detect compromised hosts. 
Although $DS_p$ accounts for 10\% of the total traffic obtained, this dataset is still larger than most datasets used in previous studies. 
\url{joinsanjose[.]com} queries are not included with this dataset.

\textbf{Real-world dataset.} The second dataset, denoted as $DS_r$, consists of 255 billion DNS queries collected in a similar fashion to the collection process of $DS_f$ and is provided by the same source.
The queries were collected over the course of 21 days, beginning on February 28, 2023.
Accordingly, the number of queries per hour is 507 million; to the best of our knowledge, this is the largest dataset ever used to evaluate DNS exfiltration detection methods. 
This dataset will be used in our real-world evaluation of \algoName{} and the compared methods in Section~\ref{sec:realworldeval}.

\textbf{Public dataset.} We perform our evaluation on a third, publicly available dataset~\cite{Ziza2022DNSExfiltrationDataset}. 
This dataset was constructed by collecting more than 35M DNS queries from an Internet service provider's (ISP) DNS server over the course of 26 hours. Accordingly, the average number of queries per hour is 1.9 million. The dataset also contains exfiltration queries to three distinct domains. The queries are generated by the Iodine~\cite{iodine} and DNSExfiltrator tools\cite{dnsexfiltrator}, both freely available on GitHub. For the rest of the paper the public dataset is denoted as $ZIZA$.



A summary of the datasets used in this study is provided in Table~\ref{tab:dataset_table}


\subsection{Parameter tuning \label{subsec:tuning}}


\begin{figure}[h]
    \centering
    \subfloat[False positive queries]{\includegraphics[width = 0.8\columnwidth]{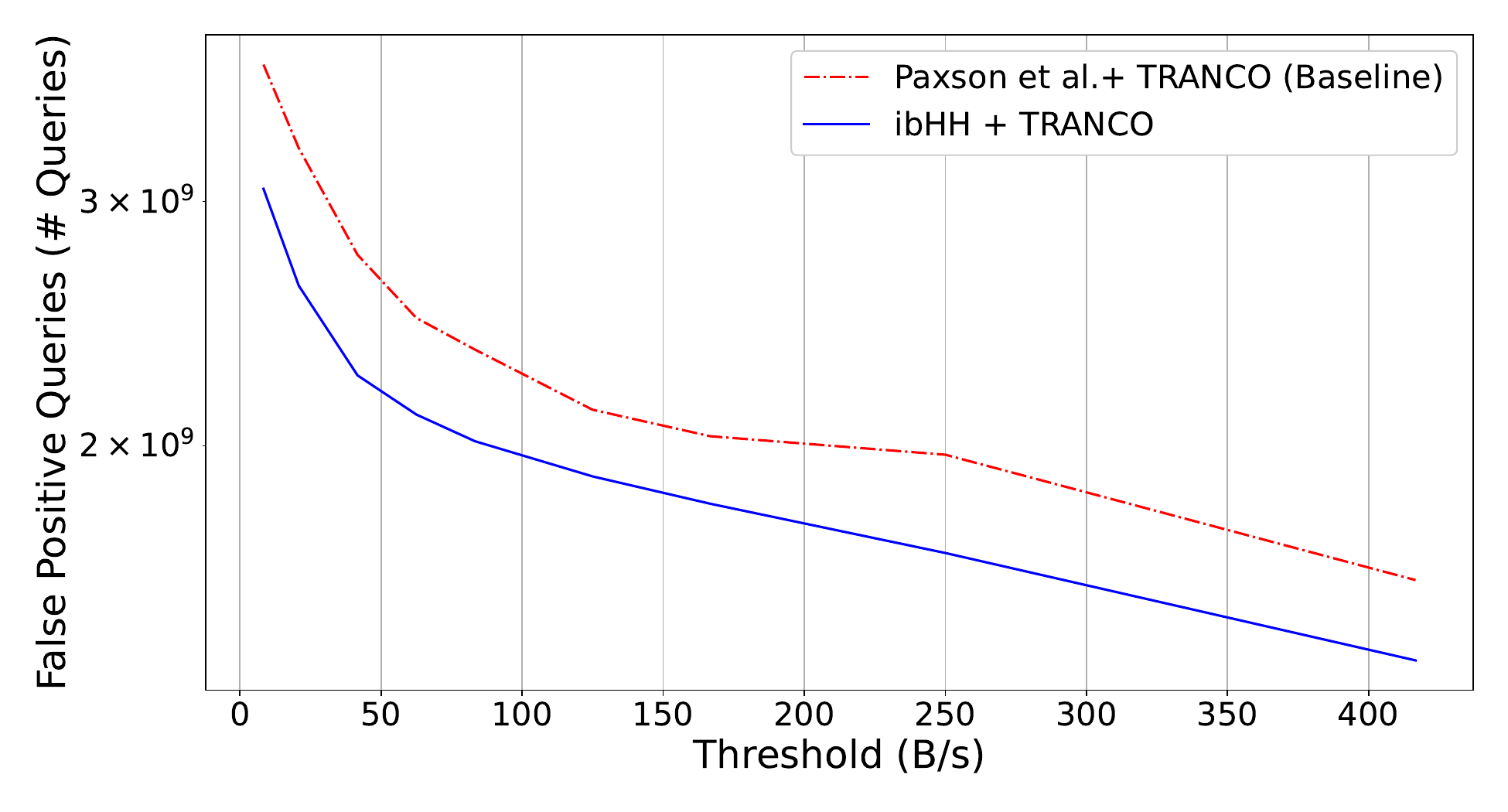}} 
    \\
    \subfloat[False positive domains ]{\includegraphics[width = 0.8\columnwidth]{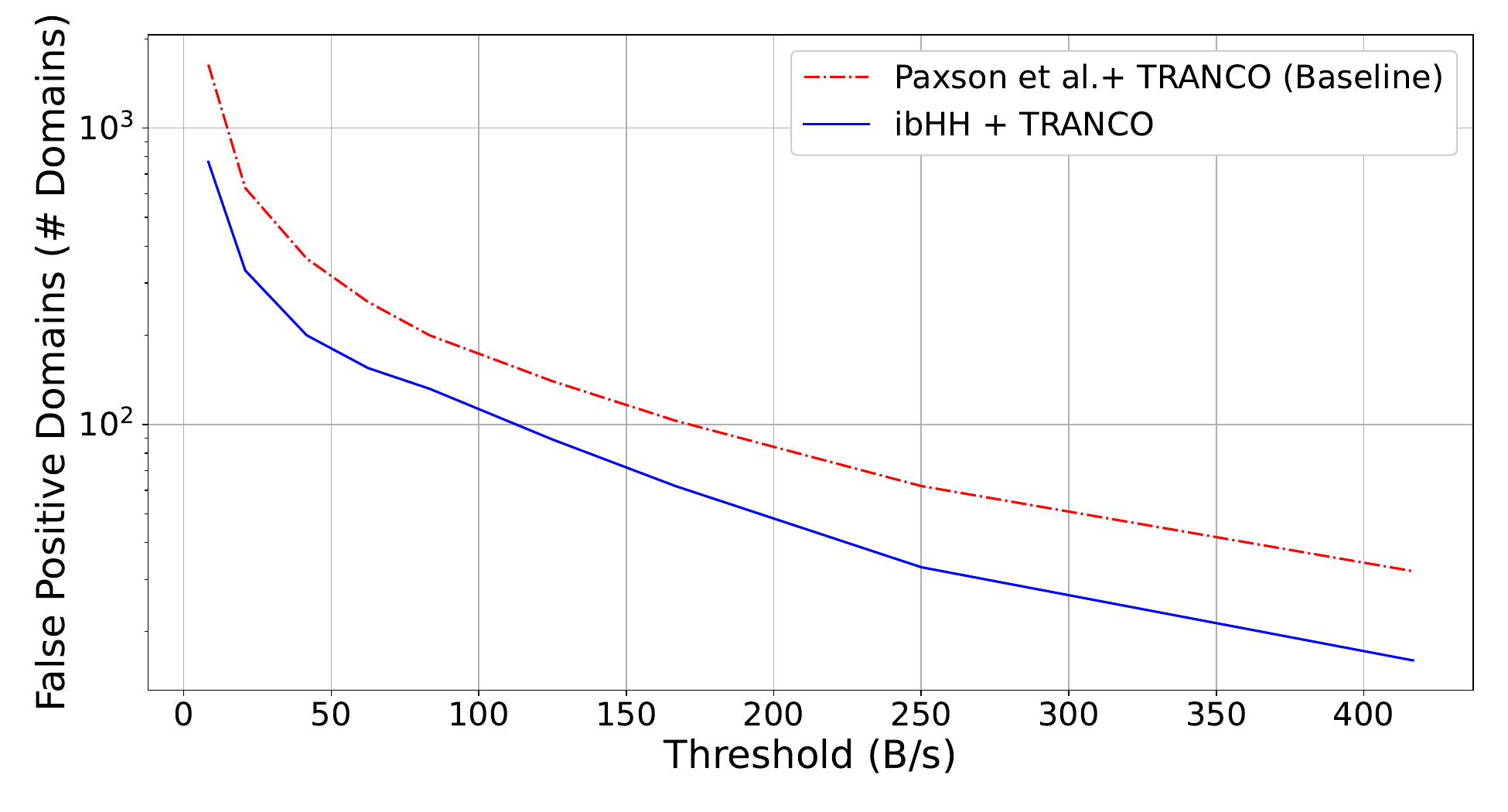}}
    \caption{Parameter tuning with TRANCO allowlist}
    \label{fig:global-fp-tranco}
\end{figure}

The objective of this tuning phase is to find the lowest detection threshold that produces a practical number of false positive domains. 
While this definition may vary between different enterprises, in this study we consider 15 false alerts per week (just over two alerts per day) to be practical. 

The detection threshold was tuned between 0 bytes/sec (B/s) and 400 B/s. 
As a baseline, we chose the method of Paxson et al.~\cite{paxson2013practical}, a similar information-based threshold detection method, which is the state-of-the-art for such methods. 
\algoName's cache size was fixed at 1,000, and the reset interval was fixed at 120 seconds.

\begin{figure}[h]
    \centering
    \subfloat[False positive queries]{\includegraphics[width = 0.8\columnwidth]{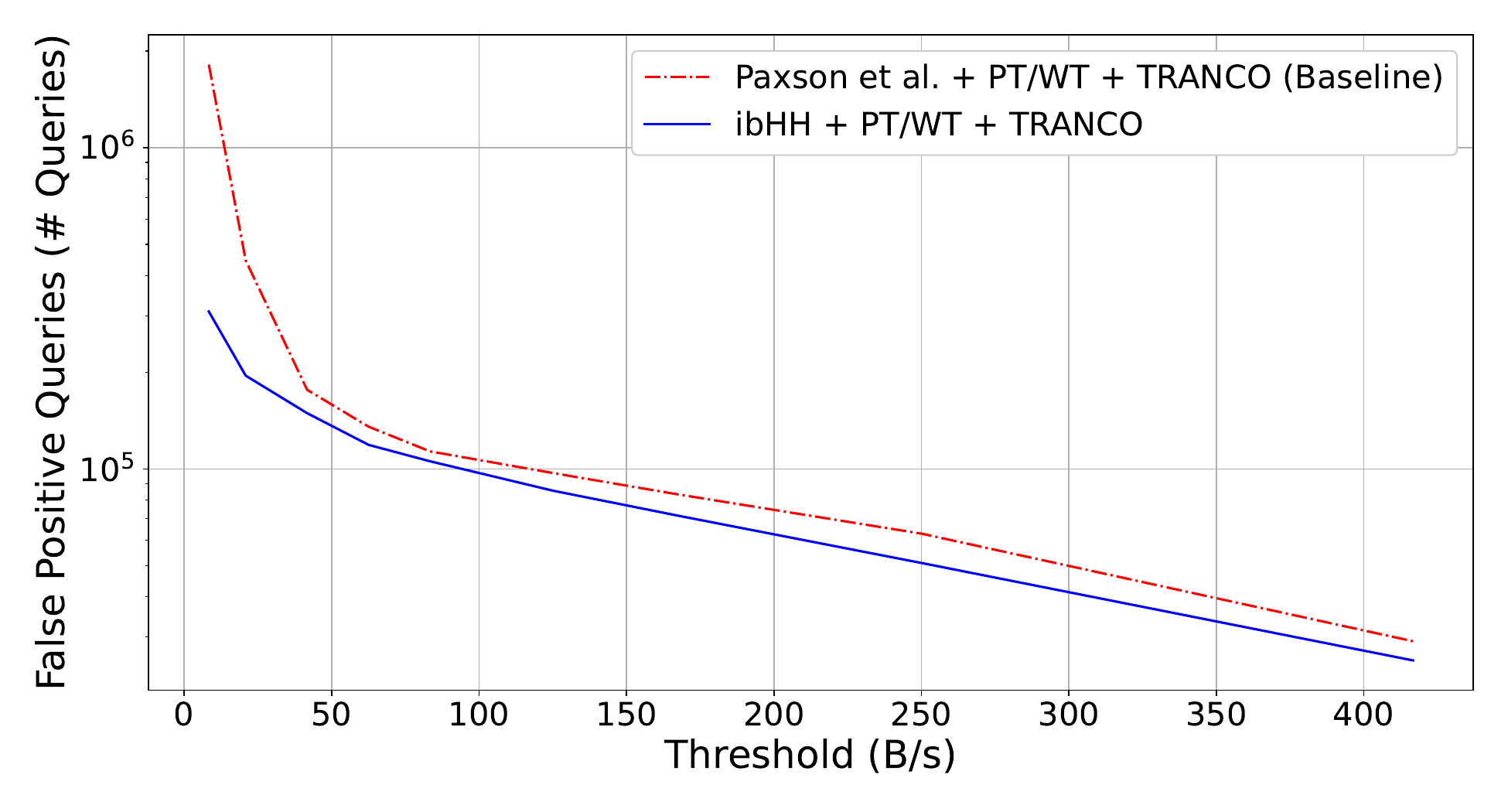}} 
    \\
    \subfloat[False positive domains ]{\includegraphics[width = 0.8\columnwidth]{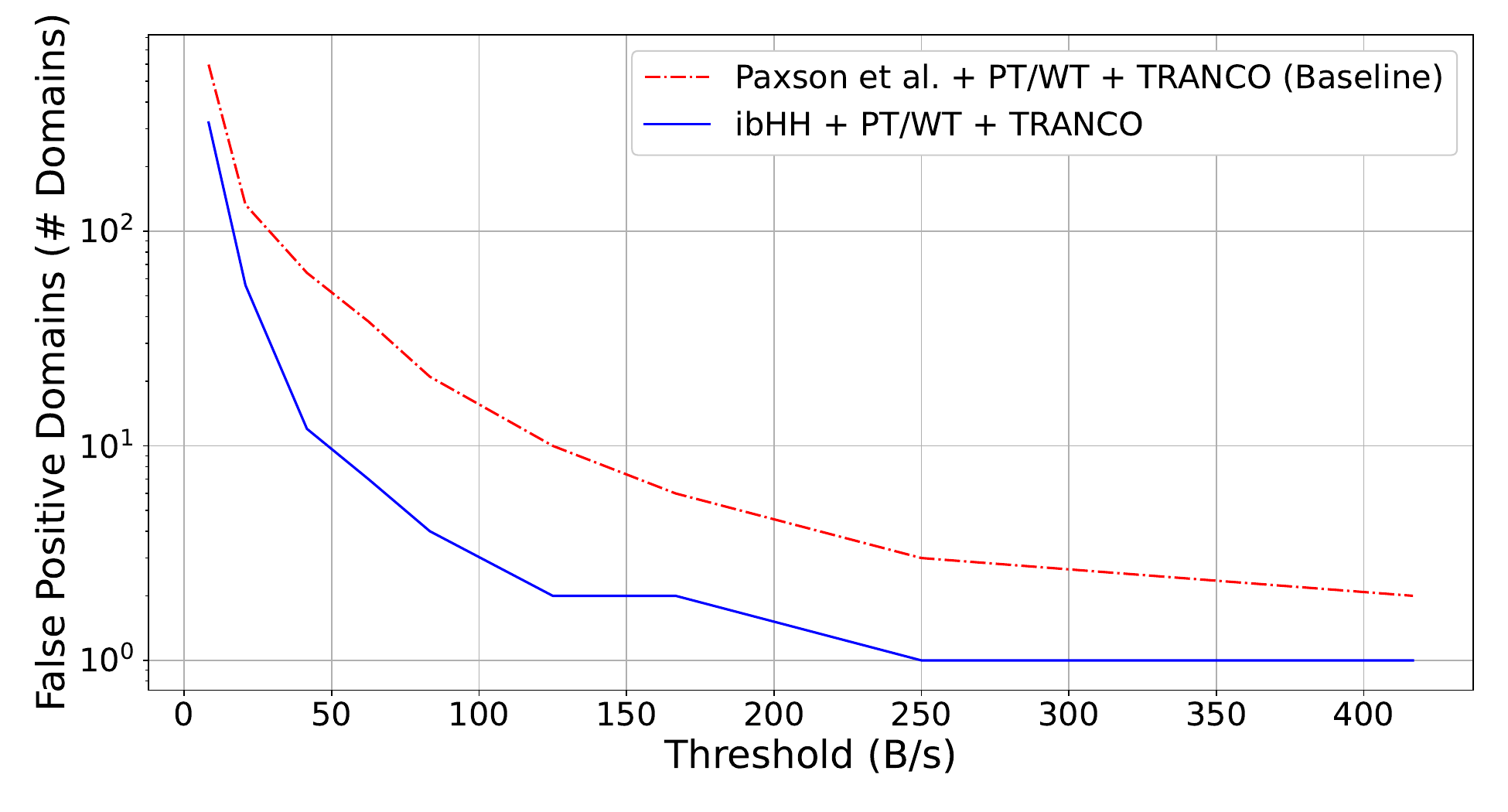}}
    \caption{Parameter tuning with TRANCO and peacetime allowlist.}
    \label{fig:global-fp-pt-tranco}
\end{figure}

In addition, we examine the effect of using the allowlisting methods we described in Section~\ref{sec:whitelist}.
To do so, we generate a PT allowlist for both \algoName{} and the method of Paxson et al. method using data from the first day.
As can be seen in Figures ~\ref{fig:global-fp} ~\ref{fig:global-fp-pt-tranco}, \algoName{} consistently produces less false positive detections than the method of Paxson et al., across all allowlist combinations.
\algoName{} with the TRANCO and PT/WT allowlists obtains a total of 10 false alerts over all 753 organizations, but it also results in a high threshold of 250 B/s. In Figure ~\ref{fig:global-fp-pt-tranco}, it can be seen that a much lower threshold of 15 B/s results in about 80 alerts over the course of seven days (an average of 0.015 alerts per day per organization), which is an acceptable alert rate for many organizations.
We can also see an exponential growth in the number of alerted domains in the lower threshold values.

\subsection{Analysis of alerted domains \label{subsec:fp-analysis}}

Based on the results presented in Section~\ref{subsec:tuning}, we manually inspect the domains for which \algoName{} raised an alert (see Table~\ref{tab:fp-analysis} which summarizes our findings).
This analysis is performed on the positive alerts produced by the \algoName{} + TRANCO + PT/WT with a detection threshold of 250 B/s configuration. 
As noted, 10 domains out of 43 million unique domains in the dataset (0.00002\% of the unique domains in the dataset) were marked as suspected for DNS exfiltration. 
Out of the 10 domains, six domains are registered and operated by known security vendors.
\url{sophosxl[.]com}, \url{appsechcl[.]com}, \url{barracudabrts[.]com}, \url{dnsbl[.]org}, and \url{softsqr[.]com} are domains used by DNS anti-malware list (DNSAML) service providers~\cite{nadler2022vulnerability}.
\url{cnr[.]io} is a domain used for honeypot~\cite{mairh2011honeypot} services.
\new{Security vendors' AV clients send DNS requests with their current signature ruleset version or suspicious file hashes encoded within the DNS request subdomains, which results in a high number of subdomains ~\cite{nadler2022vulnerability}.}
Three domains, \url{pldtgroup[.]net}, \url{cnsevrx[.]com}, and \url{kfcmsp[.]com}, are domains associated with a large number of unique (and occasionally, long) subdomains; therefore, the method marked them as suspected DNS exfiltration domains. We looked these domains up with the WHOIS ~\cite{rfc3912} protocol (a query/response protocol that is widely used to obtain information about registered domain names). All three domains were registered at least five years ago (which considerably lowers the likeliness that they have been used for DNS exfiltration), and \url{pldtgroup[.]net} is registered to the Philippine Long Distance Telephone Company, a reputable company. We thus conclude that these cases are false positive alerts.
The last alerted domain, \url{joinsanjose[.]com}, was also classified as DNS exfiltration by the CDN provider's algorithm. We were not able to conclude its maliciousness, and therefore we classify it as \textbf{unknown}, until further investigation is done.

\begin{table*}
\footnotesize
\centering
\begin{tabular}{p{2cm}| p{2.1cm}| p{2.1cm}| p{2.1cm}| p{2.4cm} p{2.4cm}| p{1.9cm}}
\toprule
 Detection window size (sec) &  Cache size (\# entries) &  Number of alerts (\# domains) &  Number of alerts with PT allowlist &  Number of alerts with TRANCO allowlist &  Number of alerts with both allowlists &  Number of true positiveS \\
\midrule
   120 &         100 &          34 &              8 &                 15 &                3 &       3 \\
              120 &        1000 &          41 &             11 &                 19 &                4 &       3 \\
              120 &       10000 &          41 &             11 &                 19 &                4 &       3 \\
              600 &         100 &          23 &              4 &                  6 &                3 &       3 \\
              600 &        1000 &          26 &              4 &                  6 &                3 &       3 \\
              600 &       10000 &          26 &              4 &                  6 &                3 &       3 \\
             1800 &         100 &          13 &              3 &                  3 &                2 &       2 \\
             1800 &        1000 &          14 &              4 &                  4 &                3 &       3 \\
             1800 &       10000 &          14 &              4 &                  4 &                3 &       3 \\
\bottomrule
\end{tabular}
\caption{Sensitivity analysis for \algoName{}.}
\label{tab:wtf}
\end{table*}

\begin{table}
\centering
\begin{adjustbox}{width={0.5\textwidth}}
\begin{tabular}{c|c|c|l}
\hline
Domain    &  Category  & Frequency   & Subdomain Example   \\ 
\hline
cnr[.]io            & \begin{tabular}[c]{@{}l@{}}Honeypot\\Service\end{tabular} & 1 out of 10 & \begin{tabular}[c]{@{}l@{}}7zn28g\\.2.592\\.ZPQXSURLT7IU5YQFCOS2S76N\\GVNZHVA2MSZE6JIBXZ2B3EES2BORGPHCI2G6FDC\\.KQRC5YIEIXH5UBLYRSTQ3C4B\\ZNNTAD7OVOQUIOD3KEP5JQLEOVGTS2F3HFS7ZGO\\.YMJULMS26QIC7RLPXGYOZHQA888888.ef584e16\end{tabular}  \\ 
\hline
dnsbl[.]org         &                                                           &             & 2FWww.wayfAIr.cOm.IN                                                                                                                                                                                                                      \\ 
\cline{1-1}\cline{4-4}
barracudabrts[.]com &                                                           &             & 01143c071e01.t-164113145.id135d030.prizelabs.com.d.bl                                                                                                                                                                                     \\ 
\cline{1-1}\cline{4-4}
sophosxl[.]com      & \begin{tabular}[c]{@{}l@{}}DNSAML ~\cite{nadler2022vulnerability} \\Services\end{tabular}  & 5 out of 10 & adfba15d2df0b4465475c420f2a2137d.sigv2.vir1                                                                                                                                                                                               \\ 
\cline{1-1}\cline{4-4}
appsechl[.]com      &                                                           &             & \begin{tabular}[c]{@{}l@{}}v3-query-13018-40e9142bf\\-95c2-41e5-87e7-d6a7c536a2e4.securityip\end{tabular}                                                                                                                                 \\ 
\cline{1-1}\cline{4-4}
softsqr[.]com       &                                                           &             & \begin{tabular}[c]{@{}l@{}}id-c55c08e136af19b29d1a33684e2f7100.\\4efeac399c6095bf867256d4529f4f1ef257\\a677c3e2e207472.fde9-0050-06cd\\-aa55d1-1BS80I039724-61cac413.f\end{tabular}                                                       \\ 
\hline
joinsanjose[.]com   &                                                           &             & \begin{tabular}[c]{@{}l@{}}519b67d9b7487348793fefdc4e9728a12e48b30af66a6ac685e\\.606.xkgoz0.hex.a10c0458d0\end{tabular}                                                                                                                   \\ 
\cline{1-1}\cline{4-4}
pldtgroup[.]net     & Unknown                                                   & 4 out of 10 & cqgvrnyjrqlaxcppf.pldt                                                                                                                                                                                                                    \\ 
\cline{1-1}\cline{4-4}
cnsevrx[.]com       &                                                           &             & tb-024753-wan2                                                                                                                                                                                                                            \\ 
\cline{1-1}\cline{4-4}
kfcmsp[.]com        &                                                           &             & D212146-wan1                              \\                                                       
\hline
\end{tabular}
\end{adjustbox}
\caption{Summary of domains for which \algoName{} raised an alert }
\label{tab:fp-analysis}
\end{table}

\subsection{Mitigating the need for data consolidation \label{subsec:mitigating}}

To provide real-time DNS exfiltration detection, the solution needs to avoid collecting and consolidating data into a single point. In order to demonstrate that \algoName{} satisfies this requirement, we compare two deployment settings: 
\begin{enumerate}[leftmargin=*]
    \item Global - All the enterprise's data is processed by a single instance of \algoName{} (simulates consolidation of data to a single point).
    \item Local - An instance of \algoName{} is allocated per enterprise (simulates deployment of \algoName{} right on the DNS gateway of the enterprise, i.e., data is not consolidated).
\end{enumerate}

\begin{figure}[htp]
    \centering
    \includegraphics[width=0.8\columnwidth]{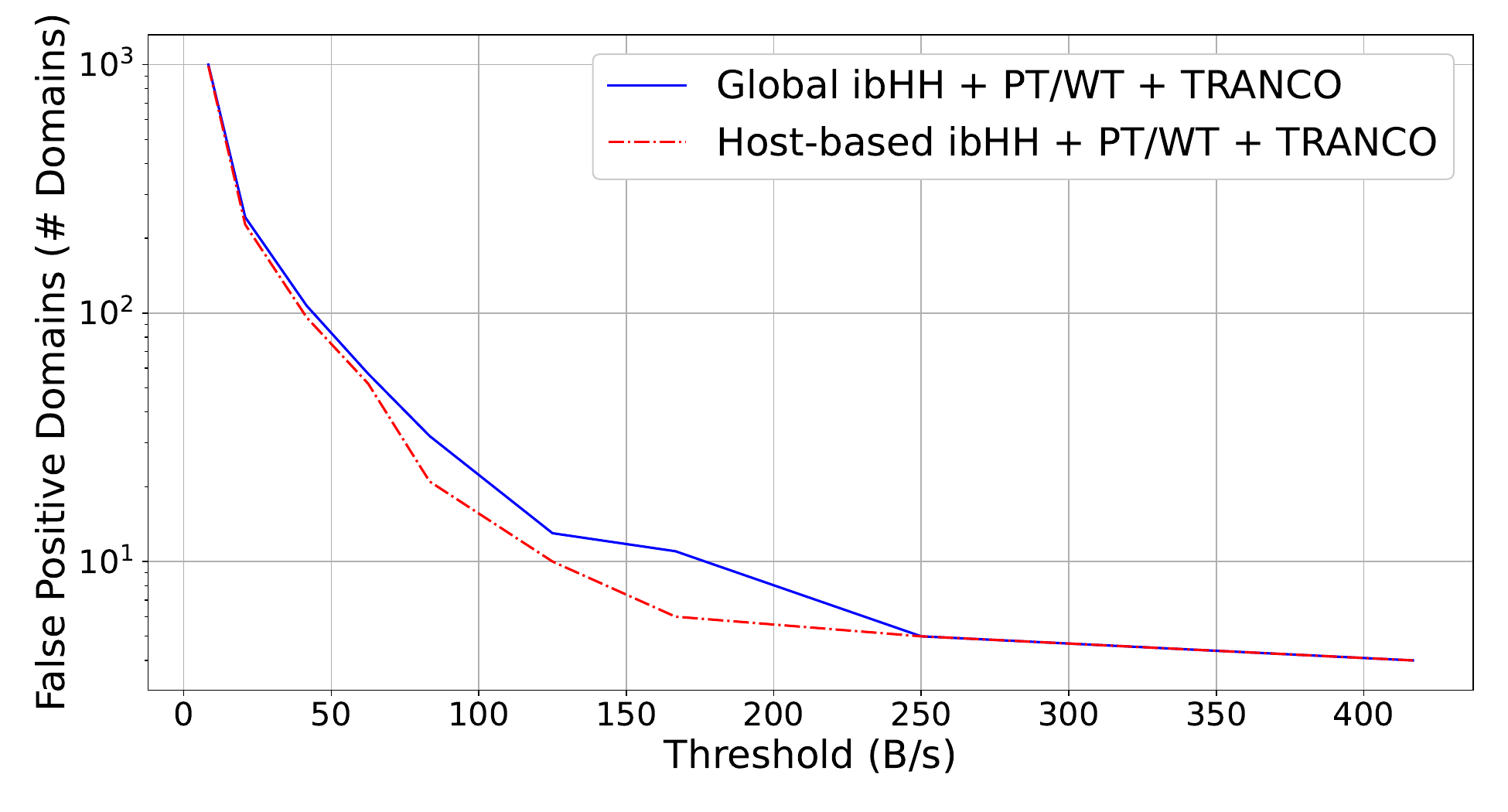}
    \caption{Global vs local false positive domains for various thresholds (\# domains).}
    \label{fig:global-vs-region-fp-domains}
\end{figure}

Similar to Section~\ref{subsec:tuning}, we tune the detection threshold and compare the number of alerted domains produced for each deployment setting. TRANCO and PT/WT allowlists are applied in both settings.
As can be seen in Figure~\ref{fig:global-vs-region-fp-domains}, the results are almost identical; thus we conclude that both deployment settings are equally viable. Therefore for \algoName{}, the data does not need to be consolidated, and it can be deployed right on the enterprise DNS gateway.



\subsection{Sensitivity analysis \label{subsec:SensitivityAnalysis}}
\new{We perform a sensitivity analysis of \algoName{}’s detection window size and cache size parameters, while setting the detection threshold at 10 B/s.
The values examined for the window size are 120, 600, and 1,800 seconds, and the examined cache size values are 100, 1,000, and 10,000. This analysis complements the detection threshold tuning step described in Section~\ref{subsec:tuning}. 
It can be seen that increasing the detection window causes a reduction in the number of alerts. 
This is expected, as a longer detection window results in a higher detection threshold for the window. 
This can have both a positive effect on the detection of false positives (reducing the number of false positive alerts) and a negative effect on the detection of true positives (when the attacker exfiltrates data in short bursts), as can be seen in the case in which the detection window is 1,800 and the cache size is 100. 
Increasing the cache size also affects the number of alerts. 
This can be attributed to the fact that in the case of a smaller cache size, a cached domain is more likely to be evicted, and thus it might not remain in the \algoName{} cache long enough to be considered an information heavy hitter. 
On the other hand, an increased cache size also means that more memory is required to store the \algoName{} cache. 
Based on this analysis, we recommend setting the detection window in the range of 120 to 600 seconds, and the cache size should be set between 1,000 and 10,000
(see Table~\ref{tab:wtf} for a summary of the results of our analysis).
}
\subsection{Compared methods\label{sec:comparedmethods}}



\subsubsection{\algoFullName{} for DNS Exfiltration Detection\label{sec:yarin}}
This is the method proposed in this paper. 
The reset interval was fixed at 120 seconds, and the cache size was set to 1,000 entries.

\subsubsection{Practical Comprehensive Bounds on Surreptitious Communication over DNS\label{sec:paxson2}}
The method of Paxson et al., which was used in the parameter tuning section~\ref{subsec:tuning}, is also used to evaluate our proposed method; it will be denoted as \emph{Paxson} for the rest of the paper. \new{\emph{Paxson} is selected as it is the SOTA information estimation-based DNS exfiltration detection technique, and the most similar to our proposed method.}

\subsubsection{Detection of Malicious and Low Throughput Data Exfiltration Over the DNS Protocol \label{sec:nadler}}
Nadler et al.~\cite{nadler2019detection} presented an unsupervised anomaly detection model based on the isolation forest algorithm~\cite{liu2008isolation} 
In this method, DNS queries are collected from recursive DNS servers at a frequency of $\lambda$ time units, and feature extraction is performed on a window size of $\lambda*n_s$, and fed to the pre-trained isolation forest model. 
\new{Despite not being a real-time solution, we chose to compare our method's performance to it, since it has the ability to detect DNS exfiltration campaigns with exfiltration rates as slow as 0.11 B/s, up to six hours after the traffic is collected. 
To the best of our knowledge, this is the SOTA in terms of near-real-time detection capabilities.}
In the remainder of the paper, we refer to this method as \emph{IF}.
We configured \emph{IF} according to the authors' recommendation, setting $\lambda = 60$ and $n_s = 6$. 

\subsubsection{Real-Time Detection of DNS Exfiltration and Tunneling from Enterprise Networks \label{sec:ahmed}}
Ahmed et al.~\cite{ahmed2019real} presented an unsupervised anomaly detection model for real-time DNS exfiltration from enterprise networks \new{based on the isloation forest algorithm}. 
\new{As noted in Section ~\ref{sec:related}, it is a true real-time method, and to the best of our knowledge, it is the SOTA real-time detection solution}.
In the remainder of the paper, this method will be referred to as \emph{RT-IF}.
We configured \emph{IF} according to the authors' recommendation, fixing the number of trees at two and limiting the tree height to 18.
All of the methods described were implemented in Python, and the experiments \new{on the $DS_p$ dataset} were performed on Azure Databricks Runtime version 10.3~\cite{azuredatabricks}. \emph{IF} and \emph{RT-IF} were both implemented using SynapseML~\cite{synapseml}. 


\subsection{Methodology \label{sec:methodology}}
$DS_{p}$ is divided into training, peacetime, and test sets. 
The training set consists of $790M$ queries from $112K$ unique hosts from the first day in the dataset. The peacetime set consists of 720M queries from the next day in the dataset. 
The rest of the dataset (six days of data) composes the test set, with a total of $3.8B$ queries from $115K$ unique hosts.
Malicious DNS exfiltration traffic is synthetically generated, similarly to previous research~\cite{nadler2019detection,ahmed2019real}. 
The attacks are generated based on three well-known DNS exfiltration tools and attacks:
\begin{enumerate}[leftmargin=*]
    \item \emph{Iodine} \cite{iodine} - Iodine is an open-source DNS tunneling tool, mainly used to bypass Wi-Fi paywalls, like the ones that can be found in hotels. This simulates high throughput DNS exfiltration campaigns.
    \item \emph{FrameworkPOS} \cite{frameworkpos} - the \emph{FrameworkPOS} malware was used in a targeted attack on the American retailer, Home Depot. Over the course of six months, the malware leaked the details of 56 million credit cards. The malware extracted credit card information from compromised machines' memory, encoded the data, and sent it to a remote server in the following format: $<encoded\_credit\_card>.domain.com$. We generate \emph{FrameworkPOS} queries at a frequency of three queries per second to simulate the original malware's throughput of 56 million credit cards in six months.
    \item \emph{Backdoor.Win32.Denis} - The Trojan malware \textbf{Backdoor.\\
    Win32.Denis} (which will be referred to as Denis for the remainder of the paper) was used in Operation Cobalt Kitty, a large-scale Asian APT \cite{denis, deniskaspersky}. 
    \emph{Denis} enables an intruder to manipulate the file system and run arbitrary commands and loadable modules. Denis uses the DNS as a bidirectional C\&C communication channel with its operator. 
    There are 16 predefined instructions to allow the C\&C operator to take control of a compromised machine. In this paper, we simulate the malware's keep-alive instructions.
    We generate the requests every 1.5 seconds, conforming to Cobalt Kitty's operation security report analysis.
\end{enumerate}

In each experiment, 1\% of the client hosts (i.e., 1,300 hosts) are sampled.
Queries are generated using one of the DNS exfiltration tools, with the sampled client hosts as the source of the DNS queries.
We evaluate the detection abilities of each of the methods based on the following metrics: number of overall hosts alerted (i.e., number of hosts suspected as being infected), hosts' $TPR$ (true positive rate, hosts which are truly infected and for which alerts are raised for), hosts' $FPR$ (hosts which are not infected but for which alerts are raised). 
Each host is infected with a random number of malicious queries in the range of 100 to 10,000, where the queries are injected at random start times in the test set. 
To identify the infected hosts, each host is associated with a distinct malicious primary domain.
We want to compare the methods' abilities to detect compromised \new{client} hosts.
Therefore the compared methods are trained with different acceptable false positive rate ($FPR$) values: 0.01 \new{(1,300 clients)}, 0.001 \new{(130 clients)}, 0.0001 \new{ (13 clients)}, 0.00001 \new{(1-2 clients)}. 
In each experiment, \emph{IF} and \emph{RT-IF} are trained by setting the isolation forest's contamination rate to be the experiment's acceptable $FPR$ value. 
For \algoName{} and Paxson, the algorithms are executed on the training dataset, and the detection threshold is tuned to be the minimum value for which the acceptable $FPR$ is achieved.
For each of the compared methods, we generate a peacetime allowlist by feeding the peacetime dataset to the trained model. The peacetime allowlist is composed of alerted domains in the peacetime dataset. 
The TRANCO allowlist is applied to all the compared methods.

\begin{table*}[h]
\begin{adjustbox}{width=\textwidth}
\begin{threeparttable}
\begin{tabular}{cccccccccccccccccccccc}
\cline{1-21}
Method &
  Dataset &
  \multicolumn{3}{c}{FPR=0.01} &
   &
   &
  \multicolumn{3}{c}{FPR=0.001} &
   &
   &
  \multicolumn{3}{c}{FPR=0.0001} &
   &
   &
  \multicolumn{3}{c}{FPR=0.00001} &
   &
   \\ \cline{1-21}
 &
   &
  $TD^1$ &
  FPR &
  TPR &
  $DER^1$ &
   &
  $TD^1$ &
  FPR &
  TPR &
  $DER^1$ &
   &
  $TD^1$ &
  FPR &
  TPR &
  $DER^1$ &
   &
  $TD^1$ &
  FPR &
  TPR &
  $DER^1$ &
   \\ \cline{1-21}
\multirow{3}{*}{ibHH} &
  $DS_{p} + I$ &
  1734 &
  \textbf{0.0037} &
  \textbf{1.0} &
  0.7 &
   &
  1420 &
  \textbf{0.001} &
  \textbf{1.0} &
  5 &
   &
  1343 &
  \textless{}0.001 &
  \textbf{1.0} &
  65 &
   &
  1300 &
  0 &
  \textbf{1.0} &
  275 &
   \\
 &
  $DS_{p} + F$ &
  1743 &
  \textbf{0.0038} &
  \textbf{1.0} &
  0.7 &
   &
  1430 &
  \textbf{0.001} &
  \textbf{1.0} &
  5 &
   &
  1298 &
  \textless{}0.001 &
  \textbf{0.98} &
  65 &
   &
  1280 &
  0 &
  \textbf{0.97} &
  275 &
   \\
 &
  $DS_{p} + D$ &
  1728 &
  \textbf{0.0037} &
  \textbf{1.0} &
  0.7 &
   &
  1417 &
  \textbf{0.001} &
  \textbf{1.0} &
  5 &
   &
  1252 &
  \textless{}0.001 &
  \textbf{0.98} &
  65 &
   &
  1214 &
  0 &
  0.92 &
  275 &
   \\
 &
  \new{$ZIZA$} &
  65 &
  \textbf{0.005 (62)} &
  \textbf{1.0 (3)} &
  0.6 &
   &
  12 &
  \textbf{0.0007 (9)} &
  1.0 (3) &
  4 &
   &
  4 &
  0.000085 (1) &
  \textbf{1.0 (3)} &
  15 &
   &
  N/A &
  N/A &
  N/A &
  N/A &
   \\ \cline{1-21}
 &
  $DS_{p} + I$ &
  3015 &
  0.007 &
  \textbf{1.0} &
   &
   &
  2132 &
  0.0012 &
  \textbf{1.0} &
   &
   &
  1342 &
  \textless{}0.001 &
  \textbf{1.0} &
   &
   &
  1300 &
  0 &
  \textbf{1.0} &
   &
   \\
IF &
  $DS_{p} + F$ &
  3015 &
  0.007 &
  0.99 &
  N/A &
   &
  2085 &
  0.0012 &
  0.96 &
  N/A &
   &
  1267 &
  \textless{}0.001 &
  \textbf{0.98} &
  N/A &
   &
  1279 &
  0 &
  \textbf{0.97} &
  N/A &
   \\
 &
  $DS_{p} + D$ &
  3015 &
  0.007 &
  0.98 &
   &
   &
  2058 &
  0.0012 &
  0.94 &
   &
   &
  1240 &
  \textless{}0.001 &
  0.97 &
   &
   &
  1183 &
  0 &
  0.91 &
   &
   \\
 &
  \new{$ZIZA$} &
  143 &
  0.012 (140) &
  \textbf{1.0 (3)} &
   &
   &
  24 &
  0.0017 (22) &
  0.67 (2) &
   &
   &
  1 &
  \textbf{0.0 (0)} &
  0.33 (1) &
   &
   &
  N/A &
  N/A &
  N/A &
   &
   \\ \cline{1-21}
 &
  $DS_{p} + I$ &
  3200 &
  0.008 &
  \textbf{1.0} &
   &
   &
  2659 &
  0.014 &
  \textbf{1.0} &
   &
   &
  1314 &
  \textless{}0.001 &
  \textbf{1.0} &
   &
   &
  1250 &
  0 &
  0.96 &
   &
   \\
RT-IF &
  $DS_{p} + F$ &
  3214 &
  0.008 &
  \textbf{1.0} &
  N/A &
   &
  2631 &
  0.014 &
  0.98 &
  N/A &
   &
  1107 &
  \textless{}0.001 &
  0.85 &
  \textbf{N/A} &
   &
  0 &
  0 &
  0 &
  N/A &
   \\
 &
  $DS_{p} + D$ &
  3170 &
  0.008 &
  0.98 &
   &
   &
  2599 &
  0.014 &
  0.95 &
   &
   &
  1039 &
  \textless{}0.001 &
  0.8 &
   &
   &
  0 &
  0 &
  0 &
   &
   \\
 &
  \new{$ZIZA$} &
  122 &
  0.01 (119) &
  \textbf{1.0 (3)} &
   &
   &
  21 &
  0.015 (19) &
  0.67 (2) &
   &
   &
  0 &
  \textbf{0.0 (0)} &
  0.0 (0) &
   &
   &
  N/A &
  N/A &
  N/A &
   &
   \\ \cline{1-21}
 &
  $DS_{p} + I$ &
  1927 &
  0.0041 &
  \textbf{1.0} &
  0.9 &
   &
  1771 &
  0.0023 &
  \textbf{1.0} &
  12 &
   &
  1314 &
  \textless{}0.001 &
  \textbf{1.0} &
  70 &
   &
  1300 &
  0 &
  \textbf{1.0} &
  300 &
   \\
Paxson &
  $DS_{p} + F$ &
  1927 &
  0.0041 &
  \textbf{1.0} &
  0.9 &
   &
  1771 &
  0.0023 &
  \textbf{1.0} &
  12 &
   &
  1249 &
  \textless{}0.001 &
  0.96 &
  70 &
   &
  1270 &
  0 &
  0.96 &
  300 &
   \\
 &
  $DS_{p} + D$ &
  1927 &
  0.0041 &
  0.98 &
  0.9 &
   &
  1771 &
  0.0023 &
  \textbf{1.0} &
  12 &
   &
  1230 &
  \textless{}0.001 &
  0.95 &
  70 &
   &
  932 &
  0 &
  0.72 &
  300 &
   \\
 &
  \new{$ZIZA$} &
  87 &
  0.0071 (84) &
  \textbf{1.0 (3)} &
  1 &
   &
  14 &
  0.0009 (11) &
  \textbf{1.0 (3)} &
  6 &
   &
  3 &
  0.000085 (1) &
  0.67 (2) &
  32 &
   &
  N/A &
  N/A &
  N/A &
  N/A &
   \\ \cline{1-21}
\end{tabular}
\scriptsize
\begin{tablenotes}
   \item[1] Total Detections (\#Distinct Hosts)
   \item[2] Detectable Exfiltration Rate (B/s)
\end{tablenotes}
\end{threeparttable}
\end{adjustbox}
\caption{Comparison of the evaluated methods based on the TPR and FPR.}
\label{tab:tpr-1}
\end{table*}

\subsection{Results \label{sec:results}}
A summary of the results is presented in Table~\ref{tab:tpr-1}, which presents the compared methods' detection abilities with different acceptable FPR values. 
For an acceptable FPR of 0.01 (1\%), \algoName{}'s detection threshold is 0.7B/s, meaning it can detect exfiltration rates as slow as 0.7 B/s while producing 1\% FP alerts on the training set. To provide context, based on our evaluation, it can be inferred that the detection time for DNS exfiltration is quick. In fact, considering the Track 2 format commonly used for credit card data, which requires at least 40 bytes of information per credit card~\cite{orange2015}, our method is capable of detecting and raising an alert within a timeframe that would typically allow for the exfiltration of at most three credit card details. This highlights \algoName{}'s effectiveness and efficiency in limiting the potential impact of data exfiltration incidents.

It should be noted that the FPR on the test set is just under 0.004, which is less than the acceptable FPR of the training set. This difference can be attributed to the allowlists, and it is observed in the rest of the methods. It should be noted that for the very low acceptable FPR value of 0.00001, $RT-IF$ is unable to detect any exfiltration events except for Iodine, while \algoName{} is able to detect the slower exfiltration rate domains used for FrameworkPOS and Denis communication.
Unsurprisingly, the lower the acceptable FPR is set, the higher the \emph{detectable exfiltration rate} ($DER$) gets. ~\algoName{} and Paxson's have a similar DER, which is to be expected, as they are both based on a similar idea of quantifying the amount of information and comparing it against a predefined threshold.

Measuring the $DER$ value of $IF$ and $RT-IF$ is tricky, since they are traffic-based and payload-based machine learning models, respectively. $IF$ collects features in a sliding window of size $\lambda$ time units, and classification is performed based on the last $n_s$ time windows. This means that exfiltration events are detected between $\lambda$ and $n_s \cdot \lambda$ time units after they occur (disregarding the time it takes to consolidate the logs into a single point, as well as the feature extraction and classification time). In the authors' recommended setup, $\lambda$ is set to 60 minutes, and $n_s$ is set to six, meaning up to six hours can pass until the detection time. Still, the theoretical detectable exfiltration rate is as slow as 0.11 B/s, making $IF$ a practical complementary solution for offline analysis. 

Because $RT-IF$ operates on on single DNS queries, it can detect DNS exfiltration events on the first packet inspected (so theoretically, it can stop DNS exfiltration by the time it starts); in practice we can see that only the model trained for an acceptable FPR of 0.01 achieves a competitive detection rate, and the method becomes less practical for acceptable FPR values of 0.001 and lower. While this method may be useful on small networks where 1\% FPR may be acceptable, it is not practical for large-scale networks like the one the dataset used in this research comes from, where an FPR of 1\% results in over 40,000 false alerts per week. 

\subsection{\new{Evaluation on the $ZIZA$ dataset} \label{sec:zizaevaluation}}
\new{We perform an additional evaluation on a public dataset, $ZIZA$ (described in Section ~\ref{subsec:dataset}). Because $ZIZA$ was collected over the course of 26 hours, we use the first 10 hours as a training set; the following two hours are used to generate the peacetime allowlist, and the final 14 hours are used as the test set. We train the different detection methods similarly to the way described in Section ~\ref{sec:methodology} and evaluate their ability to detect DNS exfiltration domains with different acceptable FPR values. The methods were compared on a single machine in order to evaluate the resource use of each method.}

\new{The results on the public dataset are similar to those obtained on $DS_p$. For an acceptable FPR of 1\%, the detection threshold of ibHH is 0.6 B/s, and it is able to detect all three DNS exfiltration domains with 62 false positive domains. While this is quite a large number, it is significantly better than that obtained by both $IF$ (140 false positive domains), $RT-IF$ (119), and $Paxson$ (87).
For the acceptable FPR of 0.01\%, $ibHH$ has only a single false positive while still detecting all the exfiltration domains. $IF$ is able to detect only one malicious domain, and Paxson detects two. $RT-IF$ is unable to detect any malicious activity in this scenario. 
See Table ~\ref{tab:tpr-1} for a summary of the results.}

\subsection{Real-world evaluation \label{sec:realworldeval}}
We perform a real-world evaluation of the compared methods on the $DS_r$ dataset described in Section~\ref{subsec:dataset}. 
We partition the dataset into training, peacetime allowlist generation, and test sets, similar to the partitioning described in Section~\ref{sec:methodology}. 
The first seven days serve as training data, followed by one day of peacetime allowlist generation, and the rest of the data is dedicated to test the trained models. 
Because this evaluation is performed on non-labeled real-world DNS queries, we cannot provide TPR and FPR estimations. 
Instead, we perform manual inspection of alerted domains and queries and determine if the alerts are true positives (TPs) or false positives (FPs) and provide these counts. 
All methods have been trained under an acceptable FPR of 0.001 (0.1\%), as it showed promising results in the synthetic dataset evaluation results for all the compared methods in Section~\ref{sec:results}. 
The detectable exfiltration rate ($DER$) \algoName{} obtained in the training phase is 6 B/s (18 credit card numbers), while \emph{Paxson}'s $DER$ is 11B/s (33 credit card numbers).

\algoName{} generated a total of 17 alerts, out of which two are true positive detections (so, 15 are false positive alerts; this is the lowers number of FP alerts for all the compared methods). 
\emph{IF} and \emph{Paxson} both successfully detected the two domains, yet with more FP alerts, while \emph{RT-IF} successfully detected only one of the domains. 
The results show that \algoName{} and \emph{IF} have a similar number of TP queries, where \emph{IF} classifies about 400 more TP queries, however with the cost of over 55,000 more FP queries than \algoName{}.
An analysis of the TP domains is provided in Section~\ref{sec:realworldtp}, and a summary table of the results is provided in Table~\ref{tab:real-world-eval}.

\subsubsection{True Positive Domain Analysis \label{sec:realworldtp}}
The first TP domain we inspect is \emph{cymulatedlp[.]com} which was detected by all models except \emph{RT-IF}. This domain name is registered by a cybersecurity vendor of a similar name and is used by enterprises to simulate exfiltration campaigns to assess data exfiltration defense mechanisms employed by the enterprises. While this is a simulated attack, we treat it as a true positive detection given the fact that the data is unlabeled. This attack consists of quite short subdomains (of length 64). The average time between two consecutive queries is approximately two seconds. This might explain why \emph{RT-IF} was not able to detect it, as it simulates a rather stealthy DNS exfiltration campaign.
The second domain, detected by all methods, is \emph{q2t[.]nl}. The data seems to be base64 encoded, with subdomain's lengths ranging between 30 and 144, and queries are sent quickly one after the other (an average of 0.01 second between consecutive queries). This domain represents a high throughput exfiltration campaign and is detected by all methods. 
We discovered with WHOIS that the domain is owned by your-freedom[.]net ~\cite{yourfreedom}, a VPN provider that supports tunneling over DNS, which further supports the classification of the domain as TP.

\begin{table}[]
\adjustbox{width=\columnwidth}{
\centering
\begin{tabular}{l|c|c|l|l|c}
\hline
\textbf{Method} &
  \multicolumn{1}{l|}{\textbf{FP Domains}} &
  \multicolumn{1}{l|}{\textbf{TP Domains}} &
  \textbf{FP Queries} &
  \textbf{TP Queries} &
  \textbf{$DER$} \\ \hline
ibHH   & \textbf{15} & \textbf{2} & 2,043  & 17,441 & 6   \\ \hline
IF     & 31          & \textbf{2} & 57,125 & 17,820 & N/A \\ \hline
RT-IF  & 20          & 1          & 5,093  & 12,391 & N/A \\ \hline
Paxson & 17          & \textbf{2} & 2,677  & 15,570 & 11  \\ \hline
\end{tabular}
}
\caption{Real-world evaluation results.}
\label{tab:real-world-eval}
\end{table}

\subsection{\new{Resource Use} \label{subsec:resource-usage}}
We evaluate the average runtime and average memory consumption of each method on a single machine with a 6 core CPU and 16GB RAM, representing a high-performance DNS server hardware specification. \emph{IF} and \emph{RT-IF} were both implemented with the scikit-learn library ~\cite{scikit-learn}. We measured the total runtime for each method to train, generate the peacetime allowlist, and classify the test dataset. The $ZIZA$ dataset was used in this evaluation. 

\algoName{} and \emph{RT-IF}, as methods with real-time capabilities, both use about 1.5MB memory, but \emph{ibHH} is significantly faster, with an average runtime of 58 seconds compared to the 857 second runtime of \emph{RT-IF}. This can be explained by the fact that $RT-IF$ needs to generate a large number of features to classify a DNS query. 

The offline methods ($Paxson$ and $IF$) have a notable disadvantage in that they need to store all the queries in a specified inspection window. This requirement leads to a considerably larger memory footprint and longer runtime, rendering them unsuitable for real-time deployment on the resolver. 
Implementing these methods on the resolver would negatively affect the rate at which the DNS resolver performs DNS resolution. 

A summary of the resource analysis is provided in Table ~\ref{tab:resource-use}.


\section{\label{sec:discussion}Discussion}

\subsection{Limitations}
\subsubsection{Allowlisting}

To distinguish between benign and malicious data exchange over DNS, we described two allowlisting methods in Section~\ref{sec:whitelist}. 
As noted in Figure~\ref{fig:global-fp}, our allowlisting approaches significantly reduce the number of FP alerts. 
Difficulty in distinguishing between malicious and benign DNS exfiltration traffic is common among DNS exfiltration detection algorithms, and allowisting methods are often employed to cope with this issue~\cite{ahmed2019real,nadler2019detection,paxson2013practical,chen2021dns}. 
Maintenance of these lists can be automated thus easing the process of incorporating them in the DNS exfiltration detection pipeline.


\subsubsection{Resilience against an aware attacker}
An aware attacker can circumvent detection by configuring malware to exfiltrate data at rates below the detection threshold. 
While this is a valid concern, we show that exfiltration campaigns as slow as \new{0.7} B/s can be detected by~\algoName{} with less than 0.04\% of benign domains misclassified. 
An IT organization that wants to detect very slow campaigns, may choose to lower the detection threshold; this will come with the cost of possibly having to deal with more false alarms.
Another challenge is the attacker's ability to use encrypted DNS requests, such as DNS over TLS~\cite{rfc7858} 
and DNS over HTTPS~\cite{rfc8484}, to evade detection. 
Enterprises can deal with this issue by blocking encrypted DNS traffic that is not resolved by an enterprise DNS resolver, as recommended by the National Security Agency~\cite{nsa2021}, and only allow encrypted DNS if it is resolved by the enterprise's internal recursive DNS resolver (which allows inspection of the raw DNS packet). 

\begin{table}[]
\adjustbox{width=\columnwidth}{
\centering
\begin{tabular}{c|c|c|c}
\hline
\multicolumn{1}{l}{\textbf{Method}} & \multicolumn{1}{l}{\textbf{Average Runtime (Seconds)}} & \multicolumn{1}{l}{\textbf{Average Memory Usage (MB)}} & \multicolumn{1}{l}{\textbf{Average Number of Queries per Second}} \\ \hline
ibHH   & \textbf{58} & \textbf{1.6} & \textbf{603,448} \\ \hline
IF     & 2,738       & 102 &  12,783      \\ \hline
RT-IF  & 857         & \textbf{1.5} & 40,840 \\ \hline
Paxson & 3,642       & 237     & 9,610     \\ \hline
\end{tabular}
}
\footnotesize
\caption{Comparison of the evaluated methods based on runtime and memory consumption, on the ZIZA dataset}
\label{tab:resource-use}
\end{table}

While \algoName{} primarily focuses on detecting DNS exfiltration through the query name, it is important to note that attackers can use other information vectors like query type and timing, thus avoid detection by \algoName{}. 
However, these vectors have limitations, such as restricted information capacity and vulnerability to inaccuracies. 
Despite these alternatives, the query name remains the most commonly exploited vector. 
By effectively detecting information conveyed through the query name, \algoName{} serves as a valuable defense against DNS exfiltration.

To the best of our knowledge, the query name has been the primary (or only) information vector utilized by all publicly available DNS exfiltration tools and known DNS exfiltration campaigns.
The method of Paxson et al. is designed to detect DNS exfiltration events regardless of the information vector, which is a strength of that approach.

Another way the attacker can try to bypass \algoName{} is to break the exfiltrated data down into single characters queries (e.g., instead of sending ``exfiltration.domain.com," the attacker would send ``e.domain.com," 
``x.domain.com," ... ,
``n.domain.com"). 
Because \algoName{} only accounts for distinct subdomains, it might miss this exfiltration scenario. However, it should be noted that this approach also results in a significantly lower rate of data exfiltration. 
In addition, DNS resolvers have a cache structure~\cite{rfc1034} that stores resolved DNS queries for a limited time. 
That can be problematic for the attacker, because subsequent requests for the same query will be served from the cache instead of being sent to the authoritative DNS server. 
Finally, the number of requests required to exfiltrate a given message increases linearly based on the message size. 
This increases the risk of DNS queries failing to reach the attacker because of DNS throttling, which is widely employed on public DNS servers~\cite{googlepublicdns,awspublicdns} (and can easily be implemented on internal enterprise DNS resolvers).


\subsection{\new{Wildcard subdomain resolution} \label{subsec:wildcard}}

\new{There are multiple services that use subdomains to host multiple services or deliver user-generated content (UGC). 
Notable examples include dropbox.com and googledocs.com, which organize their content under different subdomains and URLs for better isolation and network load distribution.
These UGC services are sometimes incorrectly classified as DNS exfiltration despite being legitimate due to their extensive use of unique subdomains, as reported by~\cite{nadler2022vulnerability}.
This is a limitation of all existing methods given their inherent inability to distinguish between legitimate and malicious cases of DNS exfiltration, and it also applies to our proposed method, which attempts to overcome this limitation by using allowlists.
This situation is suboptimal but arguably acceptable, since the rate of false alerts reported in our real-world, representative dataset of 750 organizations indicates there are, on average, less than 0.1 cases like this weekly per organization.}

\subsection{\new{Deployment considerations} \label{subsec:deployment-considerations}}
\new{Given the low time and memory complexity of \algoName{} (theoretically proven in Section~\ref{subsec:timespaceanalysis} and shown in practice in Section~\ref{subsec:resource-usage}), an organization may benefit from the deployment of multiple \algoName{} instances with different threshold values in order to cover different potential data exfiltration attacks over DNS and improve performance. This approach is also aligned with our evaluation results (see Section~\ref{sec:results}), where we present different models with different detection thresholds.}

\section{Conclusions and Future Work}
In this work, we present \algoName{}, a simple yet effective method capable of both detecting DNS exfiltration events in real time, by estimating the amount of unique information conveyed to registered domains through query subdomains, and providing explainable results (an alert is raised only if the suspected exfiltration rate exceeds a predefined threshold).

We perform an extremely comprehensive evaluation, comparing the proposed method's performance to that of prominent state-of-the-art methods, including the state-of-the-art real-time machine learning based solution that \algoName{} was shown to outperform. 
In the future, we plan to adapt \algoName{} for the detection of cross-domain exfiltration events, for example, by changing the information quantification so that it is per source user IP instead of per target registered domain. 
We also plan to explore a possible variation of \algoName{} capable of detecting other information vectors used for DNS exfiltration (such as the data exfiltration based on the query type field), as well as consider the DNS response (which can help in the detection of bidirectional communication). We also plan to deploy \algoName{} on DNS resolvers and evaluate its performance on online DNS query streams.


\bibliographystyle{IEEEtran}
\bibliography{mybibliography}

\end{document}